\documentclass[prd,showpacs,showkeys,nofootinbib]{revtex4}

\usepackage{longtable,graphicx,natbib}

\begin{document}

\title{Propagation equations for deformable test bodies with microstructure in extended theories of gravity}

\author{Dirk Puetzfeld}
\email{dirk.puetzfeld@astro.uio.no}
\homepage{http://www.thp.uni-koeln.de/~dp}
\affiliation{Institute of Theoretical Astrophysics, University of Oslo, P.O.\ Box 1029, 0315 Oslo, Norway}

\author{Yuri N. Obukhov}
\email{yo@thp.uni-koeln.de}
\affiliation{Institute for Theoretical Physics, University of Cologne, Z\"ulpicher Stra\ss e 77, 50937 K\"oln, Germany}
\altaffiliation[Also at ]{Department of Theoretical Physics, Moscow State University, 117234 Moscow, Russia}

\date{ \today}

\begin{abstract}
We derive the equations of motion in metric-affine gravity by making use of the conservation laws obtained from Noether's theorem. The results are given in the form of propagation equations for the multipole decomposition of the matter sources in metric-affine gravity, i.e., the canonical energy-momentum current and the hypermomentum current. In particular, the propagation equations allow for a derivation of the equations of motion of test particles in this generalized gravity theory, and allow for direct identification of the couplings between the matter currents and the gauge gravitational field strengths of the theory, namely, the curvature, the torsion, and the nonmetricity. We demonstrate that the possible non-Riemannian spacetime geometry can only be detected with the help of the test bodies that are formed of matter with microstructure. Ordinary gravitating matter, i.e., matter without microscopic internal degrees of freedom, can probe only the Riemannian spacetime geometry. Thereby, we generalize previous results of general relativity and Poincar\'e gauge theory.
\end{abstract}


\maketitle


\section{Introduction}\label{introduction_sec}

The relation between the field equations and the equations of motion within nonlinear gravitational theories has been subject to many works. The intimate link between these equations is one of the features of general relativity which distinguishes it from many other physical theories. The fact that, in contrast to linear field theories, the equations of motion need not to be postulated separately, but can be derived from the field equations, has been investigated shortly after the proposal of the theory. From a conceptual standpoint the derivability of the equations of motion is a very satisfactory result, since it reduces the number of additional assumptions in the theory.\footnote{The following german quotes are taken from \cite{EinsteinGrommer:1927} (translation by the authors): 
\begin{itemize}
\item ``[$\dots $] \textit{Es sieht daher so aus, wie wenn die allgemeine Relativit\"{a}tstheorie jenen \"{a}rgerlichen Dualismus bereits siegreich \"{u}berwunden h\"{a}tte. }[$\dots $]'',\\
``[$\dots $] \textit{It looks like the general theory of relativity has victoriously overcome this annoying dualism. }[$\dots $]''.
\item ``[$\dots $] \textit{Der hier erzielte Fortschritt liegt aber darin, da\ss\ zum ersten Male gezeigt ist, da\ss\ eine Feldtheorie eine Theorie des mechanischen Verhaltens von Diskontinuit\"{a}ten in sich enthalten kann. }[$\dots $] '',\\
``[$\dots $] \textit{The progress achieved in this work is that for the first time we have shown that a field theory can contain the theory of the mechanical behavior of discontinuities. }[$\dots $]''.\\
\end{itemize}} The earliest accounts of this feature of general relativity can be found in the works of \citet{Weyl:1923}, \citet{Eddington:1924}, as well as \citet{EinsteinGrommer:1927}. Nowadays this is customarily addressed as the ``problem of motion'' in the context of general relativity and other nonlinear field theories.\footnote{A historical account of works can also be found in \cite{Scheidegger:1953,Goldberg:1962:,Havas:1986,Damour:1987}.}

One may distinguish between two conceptually different methods. Both were employed in the derivation of the equations of motion within the theory of general relativity. One of them goes back to the works of \citet{EinsteinInfeldHoffman:1938,EinsteinInfeldHoffman:1938:2} and is based on the vacuum field equations of the theory. Within this method matter is modeled in the form of singularities of the field and only the exterior of bodies is considered. The second method, usually attributed to \citet{Fock:1939}, makes use of the differential conservation laws of the theory and also allows for a consideration of the interior of material bodies. In this work we are going to utilize the latter method; i.e., we base our considerations on differential identities derived from the symmetry of the action via Noether's theorem. 

In addition, we make use of a multipole decomposition of the matter currents. This allows for a systematic study of the coupling between the matter currents and field strengths of the theory at different orders of approximation. Multipole methods have been intensively studied in the context of the problem of motion since the early work of \citet{Mathisson:1937}. In table \ref{tab_timeline_multipole_works}, we provide a corresponding chronological overview.\footnote{An extended version of this table, also including works in the post-Newtonian and post-Minkowskian context, can be found in \cite{Puetzfeld:2007}}

\begin{longtable*}{llp{9.5cm}}
\caption{\label{tab_timeline_multipole_works}Timeline of works which deal with the problem of motion and multipole approximation schemes.}\\
\hline\hline
Year&Reference&Comment\\ 
\hline
1923 & \citet{Weyl:1923}& Mentions the link between the equations of motion (EOM) and the field equations.\\
1927 & \citet{EinsteinGrommer:1927} & Show that the field equations contain the EOM in GR (for a special case).\\ 
     & \citet{Lanczos:1927} & Early investigation regarding the problem of motion, treated as boundary value problem.\\
1931 & \citet{Mathisson:1931:1,Mathisson:1931:2,Mathisson:1931:3} &Systematic account of the problem of motion in GR, one of the first authors who makes use of the $\delta$-function in this context.\\
1937 & \citet{Robertson:1937} &Test particle EOM from divergence condition.\\
     & \citet{Mathisson:1937} &Possibly the earliest work utilizing a multipole method in the derivation of the EOM.\\
1938 & \citet{EinsteinInfeldHoffman:1938,EinsteinInfeldHoffman:1938:2} & Derivation of the EOM outside of material bodies. \\ 
1939 & \citet{Fock:1939} & Systematic slow motion approximation. \\ 
1940 & \citet{Papapetrou:1940:1} & Gravitational interaction of particles using the multipole method.\\ 
1941 & \citet{Lanczos:1941:2} & Test particle EOM via Gaussian integral transformation.\\
1949 & \citet{InfeldSchild:1949} & Derive the geodesic motion of test particles for empty space. \\ 
1951 & \citet{Papapetrou:1951:3} & EOM for pole-dipole test particles in GR (see also the later work \cite{Papapetrou:Urich:1955}).\\
     & \citet{Papapetrou:1951} & Derivation of the EOM utilizing a method in the spirit of \cite{Fock:1939}.\\
1953 & \citet{Papapetrou:1953} & Review of the problem of motion in GR.\\
     & \citet{Goldberg:1953} & Relationship of EOM and covariance of a field theory.\\
1955 & \citet{Meister:Papapetrou:1955} & EOM and coordinate conditions in GR.\\
1957 & \citet{Infeld:1957} & Review of approximation methods, derives EOM using Einstein-Infeld-Hoffmann (EIH) method, relaxes harmonic coordinate condition, $\delta$-function as source. \\ 
1959 & \citet{Kerr:1959:1,Kerr:1959:2} & Systematic post-Minkowskian treatment I + II (fast motion approximation). \\ 
     & \citet{Fock:1959} & Systematic slow motion/weak field approximation. \\ 
     & \citet{Tulczyjew:1959} & Test particle EOM via a simplified version of Mathisson's method.\\
1960 & \citet{InfeldPlebanski:1960} & Review of the EIH method. \\
     & \citet{Kerr:1960} & Approximation of the quasistatic case, review of three approximations schemes. \\ 
     & \citet{Synge:1960} & Integral conservation laws, EOM for mass center, energy-momentum pseudotensor definition. \\ 
1962 & \citet{Goldberg:1962:} & Review of the problems connected with the EOM in GR and the EIH method. \\ 
     & \citet{HavasGoldberg:1962} & Derive single-pole EOM by using Mathisson's method. \\ 
     & \citet{Tulczyjew:1962} & Covariant formulation of a multipole method in GR.\\
1964 & \citet{Taub:1964} & Test particle EOM in a coordinate independent manner using Papapetrou's method.\\
     & \citet{Dixon:1964} & Covariant multipole method for extended test particles in GR.\\
     & \citet{Havas:1964:2} & Generalized version of Mathisson's method in affine spaces.\\
1969 & \citet{Madore:1969} & EOM for extended bodies using a multipole method which differs from the one of \cite{Papapetrou:1951:3}.\\ 
1970 & \citet{Dixon:1970:1,Dixon:1970:2} & Extended bodies within a multipole formalism.\\
1973 & \citet{Liebscher:1973:1,Liebscher:1973:2} & EOM for pole particles in non-Riemannian spaces using the method in \cite{Madore:1969}, see also \cite{Liebscher:1979}.\\
1974 & \citet{Papapetrou:1974} & Review of the derivation of the EOM of a single-pole test particle in GR.\\
1979 & \citet{Dixon:1979} & Review of the multipole formalism in GR in the context of extended bodies.\\
1980 & \citet{Stoeger:Yasskin:1980} & Generalization of the Papapatrou equations to Poincar\'e gauge theory.\\
     & \citet{Bailey:Israel:1980} & Multipole method for the derivation of the EOM for extended bodies. \\
1987 & \citet{Damour:1987} & EOM review. \\ 
\hline\hline
\end{longtable*}

In this paper, we work out the equations of motion within a multipole formalism for a generalized gravitational theory known as metric-affine gravity (MAG) \cite{Hehl:1995}. In the theory of general relativity, the mass, or more precisely the energy-momentum, of matter is the only physical source of the gravitational field. The energy-momentum current corresponds (via the Noether theorem) to the local translational, or the diffeomorphism, spacetime symmetry. In MAG, this symmetry is extended to the local affine group that is a semidirect product of translations times the local linear spacetime symmetry group. Correspondingly, there are additional conserved currents describing microscopic characteristics of matter that arise as physical sources of the gravitational field. In continuum mechanics \cite{Cosserat:1909,Weyssenhoff:Raabe:1947,Kroener:1958,Truesdell:Toupin:1960,Mindlin:1964,Capriz:1989}, such matter is described as a medium with microstructure. In physical terms this means that the elements of a material continuum have internal degrees of freedom such as spin, dilation, and shear. The three latter microscopic sources are represented in MAG by the irreducible parts (that correspond to the Lorentz, dilational and shear-deformational subgroups of the general linear group) of the hypermomentum current. Fluid models with microstructure were extensively studied within different gravity theories (including MAG), see, e.g., \cite{Obukhov:Korotky:1987,Kopczynski:1986,Obukhov:Piskareva:1989,Kopczynski:1990,Obukhov:Tresguerres:1993}.

The metric-affine theory naturally generalizes the Poincar\'e gravity theory \cite{Obukhov:etal:1989,Obukhov:2006} in which the mass (energy-momentum) and spin are the sources of the gravitational field. The geometry that arises on the spacetime manifold is non-Riemannian, it is known as the Riemann-Cartan geometry with curvature and torsion. In MAG, this geometrical structure is further extended to the metric-affine spacetime with curvature, torsion, and nonmetricity. The resulting general scheme of MAG embeds not only Poincar\'e gravity, but also a wide spectrum of gauge gravitational models based on the conformal, Weyl, de Sitter, and other spacetime symmetry groups (for an overview, see \cite{Hehl:1995}, for example). This fact makes the analysis of the equations of motion in MAG especially interesting, with possible direct physical applications for all the gravitational models mentioned.

The energy-momentum current and the hypermomentum current (spin + dilaton + shear charge) are the sources of the gravitational field in MAG. Accordingly, test bodies that are formed of matter with microstructure have two kinds of physical properties which determine their dynamics in a curved spacetime. The properties of the first type have {\it microscopic} origin; they arise directly from the fact that the elements of a medium have internal degrees of freedom (microstructure). The properties of the second type are essentially {\it macroscopic}; they arise from the collective dynamics of matter elements characterized by mass (energy) and momentum. More exact definitions will be given later, but the qualitative picture is as follows. The averaging of the microscopic hypermomentum current yields the integrated spin, dilaton, and shear charge of a test body. In addition, the averaging of the energy-momentum and of its multipole moments gives rise to the orbital integrated momenta. In Poincar\'e gravity, there is only one relevant first moment, namely, the orbital angular momentum. It describes the behavior of a test particle as a rigid body, i.e., its rotation. In metric-affine gravity, one finds, in addition, the orbital moments that describe deformations of body. These are the orbital dilation momentum (that describes isotropic volume expansion) and the orbital shear momentum (that determines the anisotropic deformations with fixed volume). The three together (orbital angular momentum, orbital dilation momentum, and orbital shear momentum) comprise the generalized integrated orbital momentum. In this paper, we compare the gravitational interaction of the integrated hypermomentum to that of the integrated orbital momentum of a rotating and deformable test body. Thereby, we generalize the previous analysis \cite{Stoeger:Yasskin:1980} in which the effects of the integrated spin were compared to the effects of the orbital angular momentum of a rotating rigid test body.

The paper is organized as follows. In section \ref{mag_intro_sec} we recall some basic facts about the gravity theory under consideration, namely, metric-affine gravity. This is followed by a discussion of the conservation laws within this theory in section \ref{conservation_laws_sec} which form the basis for the derivation of the equations of motion. We then work out the explicit form of the propagation equations in sections \ref{sec_propa_eq_ys} and \ref{sec_propa_eq_alt}. In section \ref{relation_momenta_sec} we provide some relations between the different definitions of momenta within the multipole formalism. We discuss our findings in section \ref{conclusions_outlook_sec} and present an outlook on the open questions within this field. Our notation and conventions are summarized in appendix \ref{general_conventions_sec}. A table with the dimensions of all quantities appearing throughout the work can be found in appendix \ref{dimension_acronyms_app}.
 
\section{Metric-affine gravity}\label{mag_intro_sec}

Metric-affine gravity represents a gauge-theoretical formulation of a theory of gravitation which is based on the general affine group $A(4,R)$, i.e., the semidirect product of the four-dimensional translation group $R^{4}$ and the general linear group $GL(4,R)$. For a review of the theory see \cite{Hehl:1995,Gronwald:Hehl:1996}, and references therein. In such a theory, besides the usual ``weak'' Newton-Einstein--type gravity, described by the metric of spacetime, additional ``strong'' gravity pieces will arise that are supposed to be mediated by additional degrees of freedom related to the independent linear connection $\Gamma _{\alpha}{}^{\beta }$. Alternatively, the strong gravity pieces can also be expressed in terms\footnote{Please see appendix \ref{general_conventions_sec} on page \pageref{general_conventions_sec} for the definitions of the objects in this section and a short summary of our conventions.} of the nonmetricity $Q_{\alpha \beta}$ and the torsion $T^{\alpha }$. The propagating modes related to the new degrees of freedom are expected to manifest themselves in the non-Riemannian pieces of the curvature $R_{\alpha}{}^{\beta}$. The existence of such modes certainly depends on the choice of the dynamical scheme, or in technical terms, on the choice of the Lagrangian. The simplest generalization of the linear Hilbert-Einstein Lagrangian leads to a model with contact interaction. However, quadratic Yang-Mills--type Lagrangians describe a wide spectrum of non-Riemannian propagating gravitational modes. This is revealed, for example, by studies of generalized gravitational waves in models with torsion \cite{Adamowicz:1980,Mueller-Hoissen:Nitsch:1983,Chen:etal:1983,Sippel:Goenner:1986,Singh:Griffiths:1990,Zhytnikov:1994,Babourova:etal:1999} and in models with torsion and nonmetricity \cite{Tucker:Wang:1995,Garcia:etal:1998,Garcia:etal:2000,Macias:etal:2000,Puetzfeld:2001,King:Vassiliev:2001,Pasic:Vassiliev:2005,Obukhov:2006:1,Baekler:etal:2006}.

In a Lagrangian framework one usually considers the geometrical ``potentials'' $($metric $g_{\alpha \beta}$, coframe 1-form $\vartheta^{\alpha}$, connection 1-form $\Gamma_{\alpha }{}^{\beta })$ to be minimally coupled to matter fields, collectively called $\psi$, such that the total Lagrangian, i.e., the geometrical and the matter part, is given by \begin{equation}
L_{\text{tot}}=L\left( g_{\alpha \beta },\vartheta ^{\alpha },Q_{\alpha \beta },
T^{\alpha },R_{\alpha }{}^{\beta }\right) +L_{\text{mat}}\left( g_{\alpha \beta },
\vartheta ^{\alpha },\psi ,D\psi \right). \label{gen_mag_lagrangian}
\end{equation}
Here $D = d + \ell^\alpha{}_\beta\,\Gamma_\alpha{}^\beta$, with $\ell^\alpha{}_\beta$ denoting the generators of the linear transformations (namely, $\delta\psi = \varepsilon^\beta{}_\alpha\,\ell^\alpha{}_\beta\,\psi$, where $\varepsilon^\beta{}_\alpha$ are the infinitesimal parameters). With the following general definitions for the gauge field momenta
\begin{equation}
M^{\alpha \beta }:=-2\frac{\partial L}{\partial Q_{\alpha \beta }},\quad 
H_{\alpha }:=-\frac{\partial L}{\partial T^{\alpha }},\text{\quad }H^{\alpha }
{}_{\beta }:=-\frac{\partial L}{\partial R_{\alpha }{}^{\beta }}, 
\label{def_gen_excitations}
\end{equation}
the field equations of metric-affine gravity take the form
\begin{eqnarray}
&\left( \delta /\delta g_{\alpha \beta }\right)&   \quad DM^{\alpha \beta }-m^{\alpha \beta } =\sigma ^{\alpha \beta },  \label{feqs_0} \\
&\left( \delta /\delta \vartheta ^{\alpha }\right)& \quad  DH_{\alpha }-E_{\alpha } =\Sigma _{\alpha },  \label{feqs_1} \\
&\left( \delta /\delta \Gamma_{\alpha}{}^{\beta }\right)&  \quad DH^{\alpha }{}_{\beta }-E^{\alpha }{}_{\beta } =\Delta ^{\alpha }{}_{\beta }, \label{feqs_2} \\
&\left( \text{matter}\right) & \quad \frac{\delta L}{\delta \psi } = 0. \label{feqs_m}
\end{eqnarray}
On the right-hand side (rhs) of the field equations we have the physical sources: the metrical energy-momentum $\sigma ^{\alpha \beta }$, the canonical energy-momentum $\Sigma_{\alpha }$, and the canonical hypermomentum $\Delta^{\alpha}{}_{\beta}$ currents of the matter fields
\begin{equation}
\sigma ^{\alpha \beta }:=2\frac{\delta L_{\text{mat}}}{\delta g_{\alpha \beta }},
\quad \Sigma _{\alpha }:=\frac{\delta L_{\text{mat}}}{\delta \vartheta ^{\alpha}},
\quad \Delta ^{\alpha }{}_{\beta }:=\frac{\delta L_{\text{mat}}}{\delta 
\Gamma _{\alpha }{}^{\beta }}. \label{def_gen_matter_currents}
\end{equation}
On the left-hand side (lhs) there are typical Yang-Mills--like terms governing the gauge gravitational fields, and the corresponding terms that describe the currents of the gauge fields themselves that arise due to the nonlinearity of the theory. The metrical energy-momentum, the canonical energy-momentum, and the canonical hypermomentum currents of the gauge gravitational fields are introduced by
\begin{equation}
m^{\alpha \beta }:=2\frac{\partial L}{\partial g_{\alpha \beta }},\quad 
E_{\alpha }:=\frac{\partial L}{\partial \vartheta ^{\alpha }},\quad 
E^{\alpha }{}_{\beta }:=\frac{\partial L}{\partial \Gamma _{\alpha }{}^{\beta }}. 
\label{def_gen_gauge_currents}
\end{equation}
MAG has a wide gauge symmetry group. With the help of the Noether theorems for the diffeomorphism symmetry and for the local linear symmetry, one can verify that [provided the matter field equations (\ref{feqs_m}) are fulfilled] the following identities hold:
\begin{eqnarray}
\Sigma_\alpha &=& e_\alpha\rfloor L_{\rm mat} - (e_\alpha\rfloor D\psi)\wedge {\frac {\partial L_{\rm mat}}{\partial D\psi}} - (e_\alpha\rfloor \psi)\wedge {\frac{\partial L_{\rm mat}}{\partial \psi}},\label{SigmaA} \\
E_\alpha &=&  e_\alpha\rfloor L + (e_\alpha\rfloor T^\beta)\wedge H_\beta + (e_\alpha\rfloor R_\beta{}^\gamma)\wedge H^\beta{}_\gamma + {\frac 12}\, (e_\alpha\rfloor Q_{\beta\gamma})\,M^{\beta\gamma},\label{Ea}\\
E^\alpha{}_\beta &=& -\,\vartheta^\alpha\wedge H_\beta - M^\alpha{}_\beta, \label{Eab}\\
\Delta^\alpha{}_\beta &=& (\ell^\alpha{}_\beta\,\psi)\wedge {\frac {\partial L_{\rm mat}}{\partial D\psi}},\label{DeltaA}\\
D\Sigma _{\alpha } &=&\left( e_{\alpha}\rfloor T^{\beta }\right) \wedge \Sigma _{\beta }-\frac{1}{2}\left( e_{\alpha }\rfloor Q_{\beta \gamma }\right) \sigma ^{\beta \gamma }+\left(e_{\alpha}\rfloor R_{\beta }{}^{\gamma }\right) \wedge \Delta ^{\beta }{}_{\gamma }, \label{gen_noether_1}\\
D\Delta ^{\alpha }{}_{\beta } &=&g_{\beta \gamma }\sigma ^{\alpha \gamma }-\vartheta ^{\alpha }\wedge \Sigma _{\beta }.  \label{gen_noether_2}
\end{eqnarray}
The gauge symmetry and the corresponding Noether identities play an essential role in MAG. The most important result is as follows: It can be shown that, by means of (\ref{Ea})-(\ref{gen_noether_2}), the field equation (\ref{feqs_0}) is redundant. It is a consequence of the two other MAG field equations (\ref{feqs_1}) and (\ref{feqs_2}) and of the Noether identities. The explanation is straightforward: One can use the local linear transformations of the frames to ``gauge away" the metric $g_{\alpha\beta}$ by making it equal to the constant Minkowski metric diag$(1,-1,-1,-1)$ everywhere on the spacetime manifold. After doing this, equation (\ref{feqs_0}) is trivially solved, and one needs to solve only the remaining equations (\ref{feqs_1}) and (\ref{feqs_2}) to determine the coframe $\vartheta^\alpha$ and connection $\Gamma_\alpha{}^\beta$. 

There are many nontrivial exact solutions for different MAG models ranging from black holes, gravitational waves, to cosmological models known in the literature. Nearly all of the corresponding references can be found in the works \cite{Hehl:1995,Hehl:Macias:1999,Puetzfeld:2005:2,Baekler:Hehl:2006}.
 
\section{Conservation laws}\label{conservation_laws_sec}

An up-to-date discussion of the conservation laws within metric-affine gravity can be found in the recent work \cite{Obukhov:Rubilar:2006}. In the following sections \ref{em_conservation_subsec}-\ref{recovering_pgt_subsec} we recall the conservation laws for the canonical energy-momentum and hypermomentum. These conservation laws serve as starting point for our subsequent derivation of the propagation equations for the multipole moments of the matter currents. In \ref{recovering_pgt_subsec} we make contact with Poincar\'e gauge theory, which represents the special case of metric-affine gravity for which the distorsion, i.e., the difference between the full and the metric-compatible connection, reduces to the antisymmetric contortion, and the hypermomentum reduces to the spin current.    

\subsection{Energy-momentum conservation}\label{em_conservation_subsec}

The Noether theorem for the diffeomorphism invariance of the matter action yields the conservation law of the energy-momentum current
\begin{equation}
{\stackrel{\{\,\}}{D}}\left(\Sigma_\alpha - \Delta^\gamma{}_\beta e_\alpha\rfloor N_\gamma{}^\beta\right) \equiv \left(e_\alpha\rfloor {\stackrel{\{\,\}}{R}}_\gamma{}^\beta - {\stackrel{\{\,\}} {\hbox{\L}}}_\alpha N_\gamma{}^\beta\right)\wedge\Delta^\gamma{}_\beta.
\label{NoeD2}
\end{equation}
Here ${\stackrel{\{\,\}}{\hbox{\L}}}_\xi = \xi\rfloor{\stackrel{\{\,\}}{D}} + {\stackrel{\{\,\}}{D}}\xi\rfloor$ is the (Riemannian) covariant Lie derivative. 

After we substitute the components from (\ref{vtac})-(\ref{delc}), we finally find the tensor form of the conservation law (\ref{NoeD2}):
\begin{equation}
{\stackrel{\{\,\}}{\nabla}}_j\left(T_i{}^j - N_{ikl}\,\Delta^{klj}\right)= \big({\stackrel{\{\,\}}{R}}_{ijkl} - {\stackrel{\{\,\}}{\nabla}}_i N_{jkl}\big)\Delta^{klj}.\label{DT1}
\end{equation}
This can be identically rewritten as
\begin{equation}
{\stackrel{\{\,\}}{\nabla}}_j\,T_i{}^j = \widehat{R}_{ijkl}\,\Delta^{klj}+ N_{ikl}\,{\stackrel{\{\,\}}{\nabla}}_j\Delta^{klj},\label{DT2}
\end{equation}
where we denoted 
\begin{equation}
\widehat{R}_{ijkl} := {\stackrel{\{\,\}}{R}}_{ijkl} - {\stackrel{\{\,\}}{\nabla}}_i N_{jkl} + {\stackrel{\{\,\}}{\nabla}}_j N_{ikl}.
\label{Rhat}
\end{equation}

\subsection{Hypermomentum conservation}\label{hm_conservation_subsec}

The Noether theorem for the local $GL(4,R)$-invariance of MAG yields (on the mass shell, i.e., when the matter satisfies the field equations): 
\begin{equation}
D\Delta^\alpha{}_\beta  + \vartheta^\alpha\wedge\Sigma_\beta - \sigma^\alpha{}_\beta = 0.\label{GL}
\end{equation}
Here the last term describes the metrical energy-momentum 4-form defined in equation (\ref{def_gen_matter_currents}). By the introduction of local coordinates for the corresponding components,
\begin{equation}
\sigma^{\alpha\beta} = t^{\alpha\beta}\,\eta,
\end{equation}
we can rewrite the Noether identity (\ref{GL}) in tensorial form:
\begin{equation}
{\stackrel{\{\,\}}{\nabla}}_j\,\Delta^{klj} - N_{ij}{}^k\Delta^{jli} + N^{jli}\Delta^k{}_{ij} + T^{lk} - t^{kl} = 0.
\label{GLc}
\end{equation}
Taking the antisymmetric part, we find:
\begin{equation}
{\stackrel{\{\,\}}{\nabla}}_j\,\Delta^{[kl]j} = N_{ij}{}^{[k}\Delta^{|j|l]i} + N^{j[k|i|}\Delta^{l]}{}_{ij} + T^{[kl]}.
\label{DDa}
\end{equation}

\subsection{Recovering Poincar\'e gauge theory}\label{recovering_pgt_subsec}

The case of the Poincar\'e gauge theory is recovered when the difference of the connections reduces to the antisymmetric contortion $N_{\alpha\beta} = K_{\alpha\beta} = K_{[\alpha\beta]}$, whereas the hypermomentum reduces to the antisymmetric spin current $\Delta_{\alpha\beta} = \tau_{\alpha\beta} = \tau_{[\alpha\beta]}$. 

With the help of (\ref{DDa}), we then immediately find
\begin{equation}
K_{ikl}\,{\stackrel{\{\,\}}{\nabla}}_j\tau^{klj} = K_{ikl}\,T^{kl}+ (K_{inl}\,K_{jk}{}^n - K_{jnl}\,K_{ik}{}^n)\,\tau^{klj}.
\end{equation}
Substituting this into (\ref{DT2}), and rearranging the rhs, we have 
\begin{equation}
{\stackrel{\{\,\}}{\nabla}}_j\,T_i{}^j = R_{ijkl}\,\tau^{klj}+ K_{ikl}\,T^{kl}.\label{DT3}
\end{equation}
Here the total Riemann-Cartan curvature is recovered in the first term on the rhs:
\begin{equation}
R_{ijkl} = {\stackrel{\{\,\}}{R}}_{ijkl} 
- {\stackrel{\{\,\}}{\nabla}}_i K_{jkl} 
+ {\stackrel{\{\,\}}{\nabla}}_j K_{ikl} 
+ K_{inl}\,K_{jk}{}^n - K_{jnl}\,K_{ik}{}^n,\label{RC}
\end{equation}
in complete agreement with (\ref{RR}). 

Now, writing down explicitly the Riemannian covariant derivative, we recast (\ref{DT3}) into 
\begin{equation}
\partial_j\left(\sqrt{-g}\,T_i{}^j\right) =  \sqrt{-g}\left(\Gamma_{ij}{}^k \,T_k{}^j + R_{ijkl}\,\tau^{klj}\right).\label{DT4}
\end{equation}
Here, the first term on the rhs contains the full Riemann-Cartan connection, $\Gamma_{ij}{}^k = {\stackrel{\{\,\}}{\Gamma}}_{ij}{}^k - K_{ij}{}^k$, cf. with (\ref{K}).

It is also possible to write the conservation law in a different form. By raising the index $i$, we then straightforwardly can recast (\ref{DT3}) into 
\begin{equation}
\partial_j\left(\sqrt{-g}\,T^{ij}\right) =  \sqrt{-g}\left[\left(K^i{}_{kl} - {\stackrel{\{\,\}}{\Gamma}}_{kl}{}^i\right)T^{kl} + R^i{}_{jkl}
\,\tau^{klj}\right].\label{DT5}
\end{equation}
Thus, equation (42) of \cite{Stoeger:Yasskin:1980} is correct, it coincides with (\ref{DT5}). However, one should be careful since the position of indices in the definitions of the connection, torsion, contortion, and curvature is {\it different} from our conventions. Note also that the spin in Yasskin and Stoeger is defined with the ${\frac12}$ factor, see their definition (8) in \cite{Stoeger:Yasskin:1980}, and compare it with our definition (\ref{Del}). It is satisfactory to see that our computations regarding the conservation laws are in complete agreement with those of Yasskin and Stoeger in \cite{Stoeger:Yasskin:1980}. 

\begin{figure}
\begin{center}
\includegraphics[width=10cm]{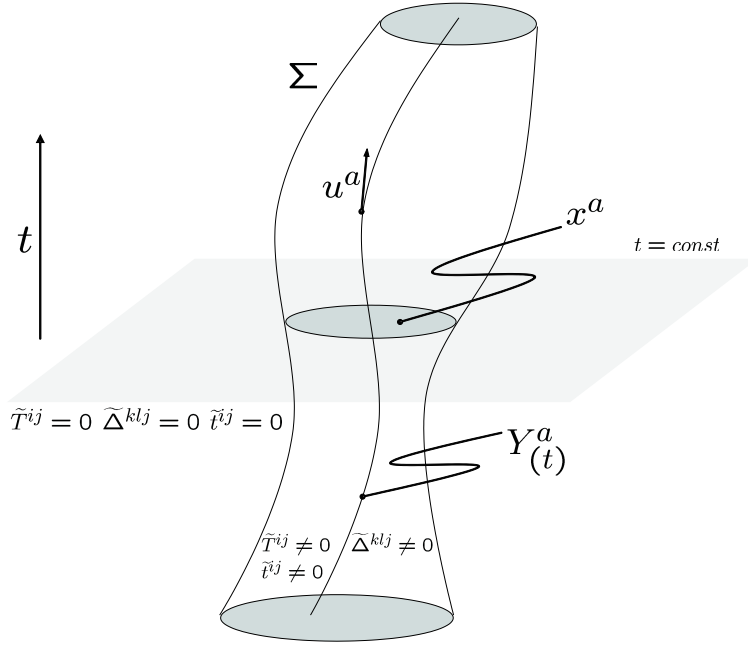}
\end{center}
\caption{\label{fig_world_tube} Sketch of the hypersurface $\Sigma$, i.e., the world tube of the test particle. A continuous curve through the tube is parametrized by $Y^a$. Coordinates within the world tube with respect to a coordinate system centered on $Y^a$ are labeled by $x^a$.}
\end{figure}

\section{Propagation equations}\label{sec_propa_eq_ys}

Let us switch to a notation which is close to the one in \cite{Stoeger:Yasskin:1980}. It turns out that (\ref{DT2}) is more appropriate to bring the energy-momentum conservation equation into a form analogous to the result (42) in \cite{Stoeger:Yasskin:1980}. By raising one index and explicitly rewriting\footnote{Remember ${\stackrel{\{\,\}}{\nabla}}_{j}\left( S^{ij}+A^{ij}\right) =\frac{1}{\sqrt{-g}}\left[ \sqrt{-g}\left(S^{ij}+A^{ij}\right) \right] _{,j}+{\stackrel{\{\,\}}{\Gamma}}_{kj}{}^{i}S^{kj},$ where $S^{ij}$ denotes the symmetric and $A^{ij}$ denotes the antisymmetric part of a quantity with two indices.} the covariant derivative in the first term of (\ref{DT2}), we obtain
\begin{equation}
\widetilde{T}^{ij}{}_{,j}{}=\widehat{R}^{i}{}_{jkl}\widetilde{\Delta }^{klj}-{\stackrel{\{\,\}}{\Gamma}}_{kj}{}^{i}\widetilde{T}^{(kj)}+N^{i}{}_{kl}{\stackrel{\{\,\}}{\nabla}}_{j}\widetilde{\Delta }^{klj}.  \label{em_conservation_ala_YS_1}
\end{equation}
Furthermore, the hypermomentum conservation equation in (\ref{GLc}) takes the form
\begin{equation}
\widetilde{\Delta }^{klj}{}_{,j}=N_{mj}{}^{k}\widetilde{\Delta }^{jlm}-{\stackrel{\{\,\}}{\Gamma}}_{mj}{}^{k}\widetilde{\Delta }^{mlj}-{\stackrel{\{\,\}}{\Gamma}}_{mj}{}^{l}\widetilde{\Delta }^{k(mj)}-N^{jlm}\widetilde{\Delta }^{k}{}_{mj}-\widetilde{T}^{lk}+\widetilde{t}^{kl}.  \label{em_conservation_ala_YS_2}
\end{equation}
By using (\ref{GLc}) in (\ref{DT2}), we can also obtain an alternative version of (\ref{em_conservation_ala_YS_1}), which has a very similar structure compared to (42) in \cite{Stoeger:Yasskin:1980}:
\begin{eqnarray}
\widetilde{T}^{ij}{}_{,j} &=&\widehat{R}^{i}{}_{jkl}\widetilde{\Delta }^{klj}+N^{i}{}_{kl}N_{aj}{}^{k}\widetilde{\Delta }^{jla}-N^{i}{}_{kl}N^{jla}\widetilde{\Delta}^{k}{}_{aj}-N^{i}{}_{kl}\widetilde{T}^{lk}-{\stackrel{\{\,\}}{\Gamma}}_{kj}{}^{i}\widetilde{T}^{(kj)}+N^{i}{}_{kl}\widetilde{t}^{kl}  \nonumber \\
\Leftrightarrow \widetilde{T}^{ij}{}_{,j} &=&R^{i}{}_{jkl}\widetilde{\Delta }^{klj}-N^{i}{}_{kl}\widetilde{T}^{lk}-{\stackrel{\{\,\}}{\Gamma}}_{kj}{}^{i}\widetilde{T}^{(kj)}+N^{i}{}_{kl}\widetilde{t}^{kl}.
\label{em_conservation_ala_YS_1_alt}
\end{eqnarray}
Note that in the last equation $R^{i}{}_{jkl}$ represents the full curvature. The structure of equation (\ref{em_conservation_ala_YS_1_alt}) is very similar to (42) in \cite{Stoeger:Yasskin:1980}. In the following we are going to derive the propagation equations for the integrated moments following from the conservation equations (\ref{em_conservation_ala_YS_2}) and (\ref{em_conservation_ala_YS_1_alt}).

\subsection{Lemma: Derivative of the integrated moments}

The following relation, cf.\ (41) in \cite{Stoeger:Yasskin:1980}, between the time derivative of the multipole expansion of a current also holds within metric-affine gravity
\begin{equation}
\frac{d}{dt}\int \left( \prod\limits_{j=1}^{n}\delta x^{b_{j}}\right) J_{A}{}^{0}=\sum\limits_{i=1}^{n}\rho ^{b_{i}}{}_{a}\int \left( \prod\limits_{j=1,\,j\neq i}^{n}\delta x^{b_{j}}\right) J_{A}{}^{a}+\int \left( \prod\limits_{j=1}^{n}\delta x^{b_{j}}\right) J_{A}{}^{a}{}_{,a}.
\label{time_deriv_current_rewritten}
\end{equation}
Here $J_{A}$ denotes the density of a matter current, in our case $\widetilde{\Delta }^{klj},$ $\widetilde{T}^{ij},$ or $\widetilde{t}^{kl}$. Additionally, we have $\delta x^{a}:=x^{a}-Y^{a}$, and $\rho^{b}{}_{a}=\delta x^{b}{}_{,a}=\delta _{a}^{b}-v^{b}\delta _{a}^{0}=\delta_{a}^{b}-\delta _{0}^{b}\delta _{a}^{0}=\delta_{\alpha}^{b}\delta_{a}^{\alpha }$ for the spatial projector. The upper-index of $J_{A}{}^{a}$ is associated with the last index of the corresponding matter current, e.g., $J_{A}{}^{0}\rightarrow \widetilde{T}^{i0}$. In (\ref{time_deriv_current_rewritten}), and in the following, integrals are taken over a 3-dimensional slice $\Sigma(t)$, at a time $t$, of the world tube of a test body. We use the condensed notation
\begin{eqnarray*}
\int\,f = \int_{\Sigma(t)}\,f(x)\,d^3x.
\end{eqnarray*} 

\subsection{Conservation equations integrated}

With the help of (\ref{time_deriv_current_rewritten}), we derive the integrated version of the conservation equations (\ref{em_conservation_ala_YS_1_alt}): 
\begin{eqnarray*}
&&\frac{d}{dt}\int \left( \prod\limits_{\alpha =1}^{n}\delta x^{b_{\alpha}}\right) \widetilde{T}^{i0}=\sum\limits_{\beta =1}^{n}\left[ \int \left(\prod\limits_{\alpha=1,\alpha \neq \beta }^{n}\delta x^{b_{\alpha }}\right) \widetilde{T}^{ib_{\beta }}-v^{b_{\beta }}\int \left( \prod\limits_{\alpha =1,\alpha \neq \beta }^{n}\delta x^{b_{\alpha }}\right) \widetilde{T}^{i0}\right]  \\
&&+\int \left( \prod\limits_{\alpha =1}^{n}\delta x^{b_{\alpha }}\right)\left( R^{i}{}_{jkl}\widetilde{\Delta }^{klj}-N^{i}{}_{kl}\widetilde{T}^{lk}-{\stackrel{\{\,\}}{\Gamma}}_{kj}{}^{i}\widetilde{T}^{(kj)}+N^{i}{}_{kl}\widetilde{t}^{kl}\right),
\end{eqnarray*}
and (\ref{em_conservation_ala_YS_2})
\begin{eqnarray*}
&&\frac{d}{dt}\int \left( \prod\limits_{\alpha =1}^{n}\delta x^{b_{\alpha}}\right) \widetilde{\Delta }^{kl0}=\sum\limits_{\beta =1}^{n}\left[ \int \left(\prod\limits_{\alpha =1,\alpha \neq \beta }^{n}\delta x^{b_{\alpha }}\right) \widetilde{\Delta }^{klb_{\beta }}-v^{b_{\beta }}\int \left( \prod\limits_{\alpha=1,\alpha \neq \beta }^{n}\delta x^{b_{\alpha }}\right) \widetilde{\Delta }^{kl0}\right]  \\
&&+\int \left( \prod\limits_{\alpha =1}^{n}\delta x^{b_{\alpha }}\right)\left( N_{mj}{}^{k}\widetilde{\Delta }^{jlm}-{\stackrel{\{\,\}}{\Gamma}}_{mj}{}^{k}\widetilde{\Delta }^{mlj}-{\stackrel{\{\,\}}{\Gamma}}_{mj}{}^{l}\widetilde{\Delta }^{k(mj)}-N^{jlm}\widetilde{\Delta}^{k}{}_{mj}-\widetilde{T}^{lk}+\widetilde{t}^{kl}\right).
\end{eqnarray*}
With the introduction of new names for the integrated moments
\begin{eqnarray}
\overline{\Delta}^{b_{1}\cdots b_{n}ijk} &:&=\int \left( \prod\limits_{\alpha =1}^{n}\delta x^{b_{\alpha }}\right) \widetilde{\Delta }^{ijk}, \nonumber \\
\overline{T}^{b_{1}\cdots b_{n}ij} &:&=\int \left( \prod\limits_{\alpha =1}^{n}\delta x^{b_{\alpha }}\right) \widetilde{T}^{ij}, \nonumber \\
\overline{t}^{b_{1}\cdots b_{n}ij} &:&=\int \left( \prod\limits_{\alpha =1}^{n}\delta x^{b_{\alpha }}\right) \widetilde{t}^{ij}, \label{int_moments_definitions}
\end{eqnarray}
the integrated conservation laws take the following form:\footnote{Note that we use an inverted circumflex, e.g., $\check{b}_\beta$, to indicate that an index is omitted from a list.}
\begin{eqnarray}
&&\frac{d}{dt}\overline{T}^{b_{1}\cdots b_{n}i0}=\sum_{\beta =1}^{n}\left(\overline{T}^{b_{1}\cdots \check{b}_{\beta }\cdots b_{n}ib_{\beta }}-v^{b_{\beta }} \overline{T}^{b_{1}\cdots \check{b}_{\beta }\cdots b_{n}i0}\right)   \nonumber \\
&&+\int \left( \prod\limits_{\alpha =1}^{n}\delta x^{b_{\alpha }}\right)
\left( R^{i}{}_{jkl}\widetilde{\Delta }^{klj}-N^{i}{}_{kl}\widetilde{T}^{lk}-{\stackrel{\{\,\}}{\Gamma}}_{kj}{}^{i}\widetilde{T}^{(kj)}+N^{i}{}_{kl}\widetilde{t}^{kl}\right),
\label{int_cons_eq_1} \\
&&\frac{d}{dt} \overline{\Delta}^{b_{1}\cdots b_{n}kl0}=\sum_{\beta =1}^{n}\left(\overline{\Delta}^{b_{1}\cdots \check{b}_{\beta }\cdots b_{n}klb_{\beta }}-v^{b_{\beta}}\overline{\Delta}^{b_{1}\cdots \check{b}_{\beta }\cdots b_{n}kl0}\right)   \nonumber \\
&&+\int \left( \prod\limits_{\alpha =1}^{n}\delta x^{b_{\alpha }}\right) \left( N_{mj}{}^{k}\widetilde{\Delta }^{jlm}-{\stackrel{\{\,\}}{\Gamma}}_{mj}{}^{k}\widetilde{\Delta }^{mlj}-{\stackrel{\{\,\}}{\Gamma}}_{mj}{}^{l}\widetilde{\Delta }^{k(mj)}-N^{jlm}\widetilde{\Delta}^{k}{}_{mj}-\widetilde{T}^{lk}+\widetilde{t}^{kl}\right) .  \label{int_cons_eq_2}
\end{eqnarray}
Equations (\ref{int_cons_eq_1}) and (\ref{int_cons_eq_2}) may be compared to (51) and (52) in \cite{Stoeger:Yasskin:1980}.

\subsection{Propagation equations for pole-dipole particles}

Let us now proceed along the lines of \cite{Stoeger:Yasskin:1980} and derive the propagation equations for pole-dipole particles by using (\ref{int_cons_eq_1}) and (\ref{int_cons_eq_2}). Here we investigate the case in which the following moments are nonvanishing: $\overline{\Delta}^{ijk}, \overline{T}^{ij}, \overline{T}^{ijk}, \overline{t}^{ij},$ and $\overline{t}^{ijk}$ -- i.e., we only take into account a pole contribution from the hypermomentum; the canonical energy-momentum and symmetric energy-momentum are considered to contribute at the pole as well as at the dipole level. This assumption is in accordance with the treatment in \cite{Stoeger:Yasskin:1980}, in which only pole contributions of the spin current were considered. Let us expand the geometrical quantities around the worldline $Y(t)$ of the test particle, cf.\ figure \ref{fig_world_tube}, into a power series in $\delta x^{a}=x^{a}-Y^{a}$. We have
\begin{eqnarray}
\left. R^{i}\,_{jkl}\right| _{x} &=&\left. R^{i}\,_{jkl}\right| _{Y}+\delta x^{a}\left. R^{i}\,_{jkl,a}\right| _{Y}+\cdots ,  \nonumber \\
\left. {\stackrel{\{\,\}}{\Gamma}}_{ij}{}^{k}\right| _{x} &=&\left. {\stackrel{\{\,\}}{\Gamma}}_{ij}{}^{k}\right| _{Y}+\delta x^{a}\left. {\stackrel{\{\,\}}{\Gamma}}_{ij}{}^{k}{}_{,a}\right| _{Y}+\cdots ,  \nonumber \\
\left. N^{i}{}_{kl}\right| _{x} &=&\left. N^{i}{}_{kl}\right| _{Y}+\delta x^{a}\left. N^{i}{}_{kl}{}_{,a}\right| _{Y}+\cdots .
\label{geom_quant_taylor}
\end{eqnarray}

The general form of the integrated conservation laws (\ref{int_cons_eq_1}) and (\ref{int_cons_eq_2})  then yields the following set of propagation equations:
\begin{eqnarray}
\frac{d}{dt} \overline{T}^{i0} &=&R^{i}\,_{jkl} \overline{\Delta}^{klj}-N^{i}{}_{kl}\overline{T}^{lk}-N^{i}{}_{kl,a} \overline{T}^{alk}-{\stackrel{\{\,\}}{\Gamma}}_{kj}{}^{i} \overline{T}^{(kj)}-{\stackrel{\{\,\}}{\Gamma}}_{kj}{}^{i}{}_{,a}\overline{T}^{a(kj)}  \nonumber \\
&&+N^{i}{}_{kl} \overline{t}^{kl}+N^{i}{}_{kl,a} \overline{t}^{akl},  \label{pd_prop_eq_1} \\
\frac{d}{dt} \overline{T}^{ai0} &=& \overline{T}^{ia}-v^{a} \overline{T}^{i0}-N^{i}{}_{kl} \overline{T}^{alk}-{\stackrel{\{\,\}}{\Gamma}}_{kj}{}^{i}\overline{T}^{a(kj)}+N^{i}{}_{kl}\overline{t}^{akl},  \label{pd_prop_eq_2} \\
0 &=&\overline{T}^{bia}+ \overline{T}^{aib}-v^{a}\overline{T}^{bi0}-v^{b}\overline{T}^{ai0},  \label{pd_prop_eq_3} \\
\frac{d}{dt}\overline{\Delta}^{kl0} &=&N_{mj}{}^{k}\overline{\Delta}^{jlm}-{\stackrel{\{\,\}}{\Gamma}}_{mj}{}^{k} \overline{\Delta}^{mlj}-{\stackrel{\{\,\}}{\Gamma}}_{mj}{}^{l}\overline{\Delta}^{k(mj)}-N^{jlm}\overline{\Delta}^{k}{}_{mj}-\overline{T}^{lk}+\overline{t}^{kl}, \label{pd_prop_eq_4} \\
0 &=&\overline{\Delta}^{kla}-v^{a}\overline{\Delta}^{kl0}-\overline{T}^{alk}+\overline{t}^{akl}.  \label{pd_prop_eq_5}
\end{eqnarray}
Here we suppressed the dependencies on the points at which certain quantities are evaluated. The set (\ref{pd_prop_eq_1})-(\ref{pd_prop_eq_5}) represents the generalization of the propagation equations (63)-(67) in \cite{Stoeger:Yasskin:1980} to metric-affine gravity.

\section{Alternative form of the propagation equations}\label{sec_propa_eq_alt}

It was pointed out by several authors, see also page 2086 in \cite{Stoeger:Yasskin:1980}, that the form of the propagation equations depends on the definition of  the integrated moments, in particular, the index position in the set of equations (\ref{int_moments_definitions}). Of course ambiguities emerge due to the integration process and the fact that the metric is not a constant. In the previous section we used the index positions which match the ones used in \cite{Stoeger:Yasskin:1980}; this allows for a direct comparison of their propagation equations with our result in metric-affine gravity. Since there is a priori no way to tell which index position in the integrated moments is the more physical one, we are also going to derive an alternative version of the propagation equations, in which integrated moments with mixed indices are used. 

From a formal standpoint, the definition with mixed indices may be favored over the definition with upper indices. Geometrically, the momentum should always be a covector, i.e., it should have a lower-index. This becomes immediately clear if we recall some basic facts from classical mechanics. The velocity is a vector (with an upper-index), $v^{a}=\dot{q}^{a}.$ Then, the momentum is, by definition, $p_{a}:=\partial L/\partial v^{a}$ -- which obviously is a covector. Hence, from this standpoint it appears plausible to consider the choice
\[
P_{a}=\int \widetilde{T}_{a}{}^{0},
\]
as definition for the momentum. In the following, we are going to work out an alternative set of propagation equations, which are based on the definitions with mixed indices. 

Once again, we start by rewriting the conservation equations for the canonical energy-momentum current (\ref{DT2}) and hypermomentum current (\ref{GLc}), which take the following form  
\begin{eqnarray}
\widetilde{T}{}_{i}{}^{j}{}_{,j} &=&R_{ijk}{}^{l}\widetilde{\Delta }^{k}{}_{l}{}^{j}+\Gamma _{ij}{}^{k}\widetilde{T}{}_{k}{}^{j}+N_{ij}{}^{k}\widetilde{t}^{j}{}_{k}, \label{em_conservation_ala_DY_1} \\
\widetilde{\Delta }^{k}{}_{l}{}^{j}{}_{,j} &=&\Gamma _{jl}{}^{m}\widetilde{\Delta }^{k}{}_{m}{}^{j}-\Gamma _{mj}{}^{k}\widetilde{\Delta }^{j}{}_{l}{}^{m}-\widetilde{T}{}_{l}{}^{k}+\widetilde{t}^{k}{}_{l}. \label{em_conservation_ala_DY_2}
\end{eqnarray}
Note that $\Gamma _{ij}{}^{k}$ represents the full connection, the last two equations should be compared to (42) and (43) in \cite{Stoeger:Yasskin:1980}. Apart from the index positions, equations (\ref{em_conservation_ala_DY_1}) and (\ref{em_conservation_ala_DY_2}) are completely equivalent to (\ref{em_conservation_ala_YS_1_alt}) and (\ref{em_conservation_ala_YS_2}).

\subsection{Conservation equations integrated}

With the help of (\ref{time_deriv_current_rewritten}), we derive the integrated version of the conservation equations (\ref{em_conservation_ala_DY_1}): 
\begin{eqnarray*}
&&\frac{d}{dt}\int \left( \prod\limits_{\alpha =1}^{n}\delta x^{b_{\alpha}}\right) \widetilde{T}_{i}{}^{0}=\sum\limits_{\beta =1}^{n}\left[ \int \left(\prod\limits_{\alpha =1,\alpha \neq \beta }^{n}\delta x^{b_{\alpha }}\right) \widetilde{T}_{i}{}^{b_{\beta }}-v^{b_{\beta }}\int \left( \prod\limits_{\alpha =1,\alpha \neq \beta }^{n}\delta x^{b_{\alpha }}\right) \widetilde{T}_{i}{}^{0}\right]  \\
&&+\int \left( \prod\limits_{\alpha =1}^{n}\delta x^{b_{\alpha }}\right) \left( R_{ijk}{}^{l}\widetilde{\Delta }^{k}{}_{l}{}^{j}+\Gamma _{ij}{}^{k} \widetilde{T}{}_{k}{}^{j}+N_{ij}{}^{k}\widetilde{t}^{j}{}_{k}\right),
\end{eqnarray*}
and (\ref{em_conservation_ala_YS_2})
\begin{eqnarray*}
&&\frac{d}{dt}\int \left( \prod\limits_{\alpha =1}^{n}\delta x^{b_{\alpha }}\right) \widetilde{\Delta }^{k}{}_{l}{}^{0}=\sum\limits_{\beta =1}^{n} \left[ \int \left(\prod\limits_{\alpha =1,\alpha \neq \beta }^{n}\delta x^{b_{\alpha }}\right) \widetilde{\Delta }^{k}{}_{l}{}^{b_{\beta }}-v^{b_{\beta }}\int \left(\prod\limits_{\alpha =1,\alpha \neq \beta}^{n}\delta x^{b_{\alpha }}\right) \widetilde{\Delta }^{k}{}_{l}{}^{0}\right]\\
&&+\int \left( \prod\limits_{\alpha =1}^{n}\delta x^{b_{\alpha }}\right) \left( \Gamma _{jl}{}^{m}\widetilde{\Delta }^{k}{}_{m}{}^{j}-\Gamma_{mj}{}^{k}\widetilde{\Delta }^{j}{}_{l}{}^{m}-\widetilde{T}{}_{l}{}^{k}+\widetilde{t}^{k}{}_{l}\right) .
\end{eqnarray*}
Now we introduce the integrated moments with mixed index positions. Note that we use an {\it underline} (lower-index position) to distinguish these definitions from the {\it overlined} (upper-index position) quantities in (\ref{int_moments_definitions})
\begin{eqnarray}
\underline{\Delta}^{b_{1}\cdots b_{n}i}{}_{j}{}^{k} &:&=\int \left( \prod\limits_{\alpha=1}^{n}\delta x^{b_{\alpha }}\right) \widetilde{\Delta }^{i}{}_{j}{}^{k}, \nonumber \\
\underline{T}^{b_{1}\cdots b_{n}}{}_{i}{}^{j} &:&=\int \left( \prod\limits_{\alpha =1}^{n}\delta x^{b_{\alpha }}\right) \widetilde{T}_{i}{}^{j},  \nonumber \\
\underline{t}^{b_{1}\cdots b_{n}i}{}_{j} &:&=\int \left( \prod\limits_{\alpha=1}^{n}\delta x^{b_{\alpha }}\right) \widetilde{t}^{i}{}_{j}.
\label{DY_int_moments_definitions}
\end{eqnarray}
With these definitions the integrated conservation laws take the following form
\begin{eqnarray}
&&\frac{d}{dt}\underline{T}^{b_{1}\cdots b_{n}}{}_{i}{}^{0}=\sum_{\beta =1}^{n}\left(\underline{T}^{b_{1}\cdots \check{b}_{\beta }\cdots b_{n}}{}_{i}{}^{b_{\beta}}-v^{b_{\beta }}\, \underline{T}^{b_{1}\cdots \check{b}_{\beta }\cdots b_{n}}{}_{i}\,^{0}\right)   \nonumber\\
&&+\int \left( \prod\limits_{\alpha =1}^{n}\delta x^{b_{\alpha }}\right) \left( R_{ijk}{}^{l}\widetilde{\Delta }^{k}{}_{l}{}^{j}+\Gamma _{ij}{}^{k} \widetilde{T}{}_{k}{}^{j}+N_{ij}{}^{k}\widetilde{t}^{j}{}_{k}\right),
\label{DY_int_cons_eq_1} \\
&&\frac{d}{dt}\underline{\Delta}^{b_{1}\cdots b_{n}k}{}_{l}{}^{0}=\sum_{\beta =1}^{n}\left(\underline{\Delta}^{b_{1}\cdots \check{b}_{\beta }\cdots b_{n}k}{}_{l}{}^{b_{\beta}}-v^{b_{\beta }}\, \underline{\Delta}^{b_{1}\cdots \check{b}_{\beta }\cdots b_{n}k}{}_{l}{}^{0}\right)   \nonumber \\
&&+\int \left( \prod\limits_{\alpha =1}^{n}\delta x^{b_{\alpha }}\right) \left( \Gamma _{jl}{}^{m}\widetilde{\Delta }^{k}{}_{m}{}^{j}-\Gamma_{mj}{}^{k}\widetilde{\Delta }^{j}{}_{l}{}^{m}-\widetilde{T}{}_{l}{}^{k}+\widetilde{t}^{k}{}_{l}\right) .  \label{DY_int_cons_eq_2}
\end{eqnarray}
Equations (\ref{DY_int_cons_eq_1}) and (\ref{DY_int_cons_eq_2}) should be compared to (\ref{int_cons_eq_1}) and (\ref{int_cons_eq_2}), as well as to equations (51) and (52) in \cite{Stoeger:Yasskin:1980}.

\subsection{Propagation equations for pole-dipole particles}

Finally, we derive the propagation equations for pole-dipole particles by using (\ref{DY_int_cons_eq_1}) and (\ref{DY_int_cons_eq_2}). Again we investigate the case in which the following moments are nonvanishing: $\underline{\Delta}^{i}{}_{j}{}^{k}, \underline{T}_{i}{}^{j}, \underline{T}^{i}{}_{j}{}^{k}, \underline{t}^{i}{}_{j},$ and $\underline{t}^{ij}{}_{k}$. The expansion of geometrical quantities around the worldline $Y(t)$ of the test particle, cf.\ figure \ref{fig_world_tube}, into a power series in $\delta x^{a}=x^{a}-Y^{a},$ reads
\begin{eqnarray}
\left. R_{ijk}{}^{l} \right| _{x} &=&\left. R_{ijk}{}^{l} \right| _{Y}+\delta x^{a}\left. R_{ijk}{}^{l}{}_{,a}\right| _{Y}+\cdots ,  \nonumber \\
\left. \Gamma _{ij}{}^{k}\right| _{x} &=&\left. \Gamma _{ij}{}^{k}\right| _{Y}+\delta x^{a}\left. \Gamma _{ij}{}^{k}{}_{,a}\right| _{Y}+\cdots , \nonumber \\
\left. N_{ij}{}^{k}\right| _{x} &=&\left. N_{ij}{}^{k}\right| _{Y}+\delta x^{a}\left. N_{ij}{}^{k}{}_{,a}\right| _{Y}+\cdots . \label{DY_geom_quant_taylor}
\end{eqnarray}

The general form of the integrated conservation laws (\ref{DY_int_cons_eq_1}) and (\ref{DY_int_cons_eq_2}) then yields the following set of propagation equations:
\begin{eqnarray}
\frac{d}{dt} \underline{T}_{i}{}^{0} &=&R_{ijk}{}^{l}\underline{\Delta}^{k}{}_{l}{}^{j}+\Gamma _{ij}{}^{k}\underline{T}_{k}{}^{j}+\Gamma _{ij}{}^{k}{}_{,a} \underline{T}^{a}{}_{k}{}^{j}+N_{ij}{}^{k} \underline{t}^{j}{}_{k}+N_{ij}\,^{k}{}_{,a}\underline{t}^{a}{}^{j}{}_{k},  \label{DY_pd_prop_eq_1} \\
\frac{d}{dt}\underline{T}^{a}{}_{i}{}^{0} &=&\underline{T}_{i}{}^{a}-v^{a}\underline{T}_{i}{}^{0}+\Gamma _{ij}{}^{k}\underline{T}^{a}{}_{k}{}^{j}+N_{ij}{}^{k} \underline{t}^{a}{}^{j}{}_{k},  \label{DY_pd_prop_eq_2} \\
0 &=& \underline{T}^{b}{}_{i}{}^{a}+\underline{T}^{a}{}_{i}{}^{b}-v^{a} \underline{T}^{b}{}_{i}{}^{0}-v^{b}\underline{T}^{a}{}_{i}{}^{0},  \label{DY_pd_prop_eq_3} \\
\frac{d}{dt}\underline{\Delta}^{k}{}_{l}{}^{0} &=&\Gamma _{jl}{}^{m}\underline{\Delta}^{k}{}_{m}{}^{j}-\Gamma _{mj}{}^{k}\underline{\Delta}^{j}{}_{l}{}^{m}-\underline{T}_{l}{}^{k}+\underline{t}^{k}{}_{l},  \label{DY_pd_prop_eq_4} \\
0 &=&\underline{\Delta}^{k}{}_{l}{}^{a}-v^{a}\underline{\Delta}^{k}{}_{l}{}^{0}-\underline{T}^{a}{}_{l}{}^{k}+\underline{t}^{ak}{}_{l}.  \label{DY_pd_prop_eq_5}
\end{eqnarray}
Again we suppressed the dependencies on the points at which certain quantities are evaluated. The set (\ref{DY_pd_prop_eq_1})-(\ref{DY_pd_prop_eq_5}) represents the generalization of the propagation equations (63)-(67) in \cite{Stoeger:Yasskin:1980} to metric-affine gravity, now with the mixed index convention. The above set should be compared to our result in (\ref{pd_prop_eq_1})-(\ref{pd_prop_eq_5}).

\subsection{Rewriting the propagation equations \`{a} la Yasskin and Stoeger}

Now let us rewrite the propagation equations of metric-affine gravity (\ref{DY_pd_prop_eq_1})-(\ref{DY_pd_prop_eq_5}) in a form which closely resembles the main theorem of Yasskin and Stoeger in Poincar\'e gauge theory, i.e., equations (53)-(58) in \cite{Stoeger:Yasskin:1980}. We start with the following identity which holds because of the definition of the projector $\rho ^{a}{}_{b}$:
\begin{equation}
\underline{\Delta}^{b}{}_{c}{}^{a}=v^{a}\underline{\Delta}^{b}{}_{c}{}^{0}+\rho ^{a}{}_{k}\underline{\Delta}^{b}{}_{c}{}^{k}.  \label{projector_identity}
\end{equation}
Using this relation, the last one of the propagation equations (\ref{DY_pd_prop_eq_5}) takes the form
\begin{equation}
\underline{T}^{alk}{}-\underline{t}^{akl}=\rho ^{a}{}_{b}\underline{\Delta}^{klb}.  \label{DY_pd_prop_eq_5_rewritten}
\end{equation}
This equation may be compared to equation (68) in \cite{Stoeger:Yasskin:1980}. Again with the help of (\ref{projector_identity}) we can rewrite (\ref{DY_pd_prop_eq_4}) as follows:
\begin{equation}
\underline{t}^{k}{}_{l}-\underline{T}_{l}{}^{k}=\nabla_{v}\underline{\Delta}^{k}{}_{l}{}^{0}+\left( \Gamma _{jm}{}^{k} \underline{\Delta}^{m}{}_{l}{}^{b}-\Gamma_{jl}{}^{m} \underline{\Delta}^{k}{}_{m}{}^{b}\right) \rho ^{j}{}_{b},  \label{DY_pd_prop_eq_4_rewritten}
\end{equation}
where 
\begin{equation}
\nabla_{v}\underline{\Delta}^{k}{}_{l}{}^{0}:=\frac{d}{dt} \underline{\Delta}^{k}{}_{l}{}^{0}+v^{m}\Gamma _{mj}{}^{k}\underline{\Delta}^{j}{}_{l}{}^{0}-v^{m}\Gamma_{ml}{}^{j} \underline{\Delta}^{k}{}_{j}{}^{0}. \label{implicit_def_fluid_deriv}
\end{equation}
Equation (\ref{DY_pd_prop_eq_4_rewritten}) should be compared to equation (69) in \cite{Stoeger:Yasskin:1980}. Proceeding along similar lines as in \cite{Papapetrou:1951:3,Stoeger:Yasskin:1980}, we are now going to cyclically permute the indices in (\ref{DY_pd_prop_eq_3}) twice, resulting in
\begin{eqnarray}
0 &=&\underline{T}^{bia}+\underline{T}^{aib}-v^{a}\underline{T}^{bi0}-v^{b}\underline{T}^{ai0},  \label{cyc_1} \\
0 &=&\underline{T}^{iab}+\underline{T}^{bai}-v^{b}\underline{T}^{ia0}-v^{i}\underline{T}^{ba0},  \label{cyc_2} \\
0 &=&\underline{T}^{abi}+\underline{T}^{iba}-v^{i}\underline{T}^{ab0}-v^{a}\underline{T}^{ib0}.  \label{cyc_3}
\end{eqnarray}
Then adding (\ref{cyc_1}) and (\ref{cyc_2}) and subtracting (\ref{cyc_3}) yields 
\begin{equation}
0=\underline{T}^{b(ai)}-\underline{T}^{a[bi]}-\underline{T}^{i[ba]}-v^{a}\underline{T}^{[bi]0}-v^{b}\underline{T}^{(ai)0}-v^{i}\underline{T}^{[ba]0}.  \label{DY_pd_prop_eq_3_rewritten}
\end{equation}
This equation should be compared to (72) in \cite{Stoeger:Yasskin:1980}. We proceed with the following identity:
\begin{equation}
2 \underline{T} ^{(ab)0}=2\underline{T}^{a[b0]}+2\underline{T}^{b[a0]}+\underline{T}^{a0b}+\underline{T}^{b0a}.  \label{sym_can_identity}
\end{equation}
Combining (\ref{sym_can_identity}) with (\ref{DY_pd_prop_eq_3}), in which we raise the index and set $i=0$, we arrive at
\begin{equation}
2\underline{T}^{(ab)0}=2\underline{T}^{a[b0]}+2\underline{T}^{b[a0]}+v^{a}\underline{T}^{b00}+v^{b}\underline{T}^{a00},  \label{intermediate_res_analogous_ys74}
\end{equation}
which should be compared to (74) in \cite{Stoeger:Yasskin:1980}. Remembering that
\begin{equation}
2\underline{T}^{[a0]0}=\underline{T}^{a00},  \label{moment_identity}
\end{equation}
which follows directly from $\delta x^{0}=0$, we can rewrite (\ref{intermediate_res_analogous_ys74}) as follows:
\begin{equation}
\underline{T}^{(ab)0}=v^{(a}\underline{\Lambda}^{b)0}+\rho ^{a}{}_{m}\underline{\Delta}^{[0b]m}+\underline{t}^{a[0b]}+\rho ^{b}{}_{m}\underline{\Delta}^{[0a]m}+\underline{t}^{b[0a]},  \label{intermediate_res_1}
\end{equation}
where we made use of (\ref{DY_pd_prop_eq_5_rewritten}) and introduced the following definition for the antisymmetric part of the integrated orbital momentum on the basis of the canonical momentum\footnote{Note that this definition corresponds to the quantity $L^{ab}$ in \cite{Stoeger:Yasskin:1980}. In this work, in contrast to \cite{Stoeger:Yasskin:1980}, we use the symbol $L^{ab}$ for the ``complete'' first moment of the integrated canonical momentum, i.e., including also the symmetric part.}
\begin{equation}
\underline{\Lambda}^{ab}:=2\underline{T}^{[ab]0}.  \label{T_L_definition}
\end{equation}
Remembering that $t^{ab}$ is a symmetric quantity, equation (\ref{intermediate_res_1}) can be rewritten as
\begin{equation}
\underline{T}^{(ab)0}=v^{(a}\underline{\Lambda}^{b)0}+\rho ^{a}{}_{m}\underline{\Delta}^{[0b]m}+\rho ^{b}{}_{m}\underline{\Delta}^{[0a]m}, \label{intermediate_res_analogous_ys76}
\end{equation}
which is analogous to equation (76) in \cite{Stoeger:Yasskin:1980}. This result can be used to rewrite (\ref{DY_pd_prop_eq_3_rewritten}), 
\begin{eqnarray}
\underline{T}^{b(ai)} &=&\underline{T}^{a[bi]}+\underline{T}^{i[ba]}+v^{a}\underline{T}^{[bi]0}+v^{i}\underline{T}^{[ba]0}  \nonumber \\
&&+v^{b}\left(\underline{T}^{a[i0]}+\underline{T}^{i[a0]}+v^{a}\underline{T}^{[i0]0}+v^{i}\underline{T}^{[a0]0}\right),
\end{eqnarray}
which resembles the first part of (77) in \cite{Stoeger:Yasskin:1980} and can finally be brought into the form
\begin{equation}
\underline{T}^{b(ai)}=\rho ^{b}{}_{m}\left( -v^{(a}\underline{\Lambda}^{i)m}+\rho ^{a}{}_{n}\underline{\Delta}^{[im]n}+\rho ^{i}{}_{n}\underline{\Delta}^{[am]n}\right), \label{intermediate_res_analogous_ys77_2nd_line}
\end{equation}
which is analogous to the second part of (77) in \cite{Stoeger:Yasskin:1980}. The last equation can be used in (\ref{DY_pd_prop_eq_5_rewritten}) to obtain
\begin{equation}
\underline{t}^{akl}=\rho ^{a}{}_{b}\left(\underline{\Delta}^{(kl)b}+v^{(l}\underline{\Lambda}^{k)b}-\rho ^{l}{}_{n}\underline{\Delta}^{[kb]n}-\rho ^{k}{}_{n}\underline{\Delta}^{[lb]n}\right).
\end{equation}
After reinsertion into (\ref{DY_pd_prop_eq_5_rewritten}) we arrive at the final result,
\begin{equation}
\underline{T}^{alk}=\rho ^{a}{}_{b}\left( \frac{1}{2}\underline{\Delta}^{lkb}+\underline{\Delta}^{klb}+v^{(l}\underline{\Lambda}^{k)b}-\rho ^{l}{}_{n}\underline{\Delta}^{[kb]n}-\rho ^{k}{}_{n}\underline{\Delta}^{[lb]n}\right),  \label{final_res_analogous_ys56}
\end{equation}
which closely resembles the form of one of the propagation equations found \cite{Stoeger:Yasskin:1980}, i.e., equation (56). With the help of (\ref{DY_pd_prop_eq_5_rewritten}), (\ref{intermediate_res_analogous_ys76}), and (\ref{final_res_analogous_ys56}), equation (\ref{DY_pd_prop_eq_2}) can now be transformed into 
\begin{eqnarray}
\underline{T}_{i}{}^{a} &=&v^{a}\underline{P}_{i}+\frac{d}{dt}\left[ \frac{1}{2}\underline{\Lambda}^{a}{}_{i}+g_{il}\left( v^{(a}\underline{\Lambda}^{l)0}+\rho^{a}{}_{m}\underline{\Delta}^{[0l]m}+\rho ^{l}{}_{m}\underline{\Delta}^{[0a]m}\right) \right]   \nonumber \\
&&-\Gamma _{ijk}\rho ^{a}{}_{b}\left( \frac{1}{2}\underline{\Delta}^{kjb}+\underline{\Delta}^{jkb}+v^{(k}\underline{\Lambda}^{j)b}-\rho ^{k}{}_{n}\underline{\Delta}^{[jb]n}-\rho ^{j}{}_{n}\underline{\Delta}^{[kb]n}\right)   \nonumber \\
&&-N_{ijk}\rho ^{a}{}_{b}\left(\underline{\Delta}^{(jk)b}+v^{(k}\underline{\Lambda}^{j)b}-\rho ^{k}{}_{n}\underline{\Delta}^{[jb]n}-\rho ^{j}{}_{n}\underline{\Delta}^{[kb]n}\right) ,  \label{final_res_analogous_ys55}
\end{eqnarray}
where we introduced $\underline{P}_{i}:=\underline{T}_{i}{}^{0}$ for the integrated 4-momentum. Equation (\ref{final_res_analogous_ys55}) is analogous to the propagation equation (55) in \cite{Stoeger:Yasskin:1980}. With the help of (\ref{final_res_analogous_ys55}) we can can bring (\ref{DY_pd_prop_eq_4_rewritten}) into the form 
\begin{eqnarray}
\nabla _{v}\underline{\Delta}^{k}{}_{l}{}^{0} &=&\underline{t}^{k}{}_{l}-v^{k}\underline{P}_{l}+\frac{d}{dt}\left[ \frac{1}{2}\underline{\Lambda}^{k}{}_{l}+g_{ln}\left( v^{(k}\underline{\Lambda}^{n)0}+\rho ^{k}{}_{m}\underline{\Delta}^{[0n]m}+\rho ^{n}{}_{m} \underline{\Delta}^{[0k]m}\right) \right]   \nonumber \\
&&-\Gamma _{ljc}\rho ^{k}{}_{b}\left( \frac{1}{2}\underline{\Delta}^{cjb}+\underline{\Delta}^{jcb}+v^{(c}\underline{\Lambda}^{j)b}-\rho^{c}{}_{n} \underline{\Delta}^{[jb]n}-\rho^{j}{}_{n} \underline{\Delta}^{[cb]n}\right)   \nonumber \\
&&-N_{ljc}\rho ^{k}{}_{b}\left(\underline{\Delta}^{(jc)b}+v^{(c}\underline{\Lambda}^{j)b}-\rho ^{c}{}_{n}\underline{\Delta}^{[jb]n}-\rho^{j}{}_{n}\underline{\Delta}^{[cb]n}\right) ,  \label{final_res_analogous_ys79}
\end{eqnarray}
which can be viewed as the analogue to (79) in \cite{Stoeger:Yasskin:1980}. Because of the different symmetries in metric-affine gravity the method used in this section, which was outlined in \cite{Stoeger:Yasskin:1980}, does not lead to a very compact form of equation (54). The last equation in the rewritten set is the one relating the time derivative of the momentum to the other matter quantities; from (\ref{DY_pd_prop_eq_1}), (\ref{projector_identity}), (\ref{DY_pd_prop_eq_4_rewritten}), and (\ref{DY_pd_prop_eq_5_rewritten}) we obtain
\begin{eqnarray}
\frac{d}{dt}\underline{P}_{i} &=&R_{ijk}{}^{l}\left( v^{j}\underline{\Delta}^{k}{}_{l}{}^{0}+\rho ^{j}{}_{n}\underline{\Delta}^{k}{}_{l}{}^{n}\right)   \nonumber \\
&&+\Gamma _{ij}{}^{k}\left[ \nabla _{v}\underline{\Delta}^{j}{}_{k}{}^{0}+\rho ^{n}{}_{l}\left( \Gamma _{nm}{}^{j}\underline{\Delta}^{m}{}_{k}{}^{l}-\Gamma _{nk}{}^{m}\underline{\Delta}^{j}{}_{m}{}^{l}\right) \right]   \nonumber \\
&&+\Gamma _{ij}{}^{k}{}_{,a}\rho ^{a}{}_{b}\underline{\Delta}^{j}{}_{k}{}^{b}+{\stackrel{\{\,\}}{\Gamma}}_{ij}{}^{k}\underline{t}^{j}{}_{k}+{\stackrel{\{\,\}}{\Gamma}}_{ij}{}^{k}{}_{,a}\underline{t}^{aj}{}_{k}.  \label{final_res_analogous_ys80}
\end{eqnarray}
This equation should be compared to (80) in \cite{Stoeger:Yasskin:1980}. We only note that an elimination of $\underline{t}^{j}{}_{k}$ and $\underline{t}^{aj}{}_{k}$ in the last two terms of (\ref{final_res_analogous_ys80}) is possible by using (\ref{DY_pd_prop_eq_5_rewritten}) and (\ref{final_res_analogous_ys55}). In the next section we work with a slightly different set of quantities, which allow for a very condensed form of the propagation equations of metric-affine gravity.

\subsection{Rewriting the propagation equations}

In this section we present a more condensed form of the propagation equations. Thereby we find a direct generalization of the main result\footnote{Please note the typo in equation (53) of \cite{Stoeger:Yasskin:1980}. Using the notation of \cite{Stoeger:Yasskin:1980} the last term in (53) should read: $\dots + \frac{1}{2} \rho^\delta{}_{\nu} N^{\beta \alpha \nu} g^{\gamma \epsilon} \nabla_\epsilon \lambda_{\alpha \beta \delta}$.} of \cite{Stoeger:Yasskin:1980}, i.e., equations (53)-(58), in the case of metric-affine gravity. 

We introduce the following notation for the integrated quantities: $\underline{P}_i := \underline{T}_i{}^0$ denotes again the integrated 4-momentum, $\underline{L}^k{}_l := \underline{T}^k{}_l{}^0$ the total orbital canonical energy-momentum and $\underline{Y}^k{}_l := \underline{\Delta}^k{}_l{}^0$ the integrated intrinsic hypermomentum. Furthermore, recalling that the hypermomentum comprises the spin, dilaton charge, and intrinsic shear, it is convenient to denote the antisymmetric part of the hypermomentum as the integrated spin $\underline{\tau}^k{}_l :=\underline{\Delta}^{[k}{}_{l]}{}^0$, whereas the trace of the hypermomentum defines the integrated dilaton charge $\underline{Z} := \underline{\Delta}^k{}_k{}^0$. 

In addition, we introduce a shorter notation for the ``convective currents," i.e., the projected quantities which we have used in previous sections and which are also used in \cite{Stoeger:Yasskin:1980}. For the intrinsic hypermomentum, we have 
\begin{eqnarray}
{\stackrel {(c)}{\underline{\Delta}}}{}^k{}_l{}^m := \underline{\Delta}^k{}_l{}^m - v^m\,\underline{\Delta}^k{}_l{}^0 \equiv \rho^m{}_n \underline{\Delta}^k{}_l{}^n, \nonumber 
\end{eqnarray}
and for the orbital canonical energy-momentum 
\begin{eqnarray}
{\stackrel {(c)}{\underline{T}}}{}^k{}_l{}^m :=\underline{T}^k{}_l{}^m - v^m\,\underline{T}^k{}_l{}^0 \equiv \rho^m{}_n \underline{T}^k{}_l{}^n . \nonumber
\end{eqnarray}
The convective spin and dilaton currents arise as the antisymmetric part and the trace of the convective current of the intrinsic hypermomentum, i.e., as 
\begin{eqnarray}
{\stackrel {(c)}{\underline{\tau}}}{}^k{}_l{}^m :=\underline{\Delta}^{[k}{}_{l]}{}^m - v^m\,\underline{\tau}^k{}_l\nonumber
\end{eqnarray}
and 
\begin{eqnarray}
{\stackrel {(c)}{\underline{Z}}}{}^k := \underline{Z}^k- v^k\,\underline{Z},\nonumber  
\end{eqnarray}
respectively (here $\underline{Z}^k := \underline{\Delta}^j{}_j{}^k$). With this notation, we recast the propagation equations 
(\ref{DY_pd_prop_eq_2})-(\ref{DY_pd_prop_eq_5}) into 
\begin{eqnarray}
\underline{T}_k{}^i &=& v^i\,\underline{P}_k + {\frac {d}{dt}}
\,\underline{L}^i{}_k - {\stackrel{\{\,\}}{\Gamma}}_{kj}{}^{l}
\,\underline{T}^i{}_l{}^j + N_{kj}{}^{l}\,{\stackrel {(c)}{\underline{\Delta}}}
{}^j{}_l{}^i,\label{DY_pd_prop_eq_2a}\\
{\stackrel {(c)}{\underline{T}}}{}^{(a}{}_i{}^{b)} &=& 0,
\label{DY_pd_prop_eq_3a}\\
\nabla_v\,\underline{Y}^i{}_k &=& -\,\underline{T}_k{}^i + \underline{t}^i{}_k
- \Gamma_{jl}{}^{i}\,{\stackrel {(c)}{\underline{\Delta}}}{}^l{}_k{}^j 
+ \Gamma_{jk}{}^{l}\,{\stackrel {(c)}{\underline{\Delta}}}{}^i{}_l{}^j,
\label{DY_pd_prop_eq_4a}\\
{\stackrel {(c)}{\underline{\Delta}}}{}^k{}_l{}^a &=& 
\underline{T}^a{}_l{}^k - \underline{t}^{ak}{}_l.\label{DY_pd_prop_eq_5a}
\end{eqnarray}
Equation (\ref{DY_pd_prop_eq_2a}) describes the canonical energy-momentum in terms of the usual combination of the ``translational" plus ``orbital" contributions (the first two terms), plus the additional contribution of the first moments. One should compare this with the alternative formula (\ref{final_res_analogous_ys55}). Equation (\ref{DY_pd_prop_eq_3a}) simply tells us that the convective current ${\stackrel {(c)}{\underline{T}}}{}^{a}{}_i{}^{b}$ is antisymmetric in the upper indices $a$ and $b$. This is a useful technical fact. The next equation (\ref{DY_pd_prop_eq_4a}) is actually an equation of motion for the intrinsic hypermomentum. Its  form closely follows the Noether conservation law of the hypermomentum, cf.\ (\ref{GL}) and (\ref{GLc}). An alternative form of such a dynamical equation for the hypermomentum is given in (\ref{final_res_analogous_ys79}). Finally, the equation (\ref{DY_pd_prop_eq_5a}) expresses the convective intrinsic hypermomentum current in terms of the first moments of the energy-momentum. 

Equations (\ref{DY_pd_prop_eq_2a})-(\ref{DY_pd_prop_eq_5a}) are easily derived from (\ref{DY_pd_prop_eq_2})-(\ref{DY_pd_prop_eq_5}), one only needs to rearrange some terms. In contrast to this, we need some additional steps to arrive at a new form of equation (\ref{DY_pd_prop_eq_1}), which represents the most interesting of the propagation equations from a physical point of view.  

We start by expanding the general connection in (\ref{DY_pd_prop_eq_1}), this yields
\begin{equation}
\frac{d}{dt}\underline{T}_{i}{}^{0}=R_{ijk}{}^{l}\underline{\Delta}^{k}{}_{l}{}^{j}+{\stackrel{\{\,\}}{\Gamma}}_{ik}{}^{l}\underline{T}_{l}{}^{k}-N_{ik}{}^{l}\left(\underline{T}_{l}{}^{k}-\underline{t}^{k}{}_{l}\right) +{\stackrel{\{\,\}}{\Gamma}}_{ik}{}^{l}{}_{,a}\, \underline{T}^{a}{}_{l}{}^{k}-N_{ik}{}^{l}{}_{,a}\, \left(\underline{T}^{a}{}_{l}{}^{k}-\underline{t}^{ak}{}_{l}\right).  \label{DY_deriv_inter_1}
\end{equation}
Furthermore, we have 
\begin{equation}
\frac{d}{dt}\left(\underline{T}_{i}{}^{0}-N_{ik}{}^{l}\underline{\Delta}^{k}{}_{l}{}^{0}\right) =\frac{d}{dt}\underline{T}_{i}{}^{0}-v^{a}N_{ik}{}^{l}{}_{,a} \, \underline{\Delta}^{k}{}_{l}{}^{0}-N_{ik}{}^{l}\frac{d}{dt}\underline{\Delta}^{k}{}_{l}{}^{0}.  \label{DY_deriv_inter_2}
\end{equation}
Insertion of (\ref{DY_pd_prop_eq_4}) and (\ref{DY_deriv_inter_1}) into (\ref{DY_deriv_inter_2}) yields
\begin{eqnarray}
\frac{d}{dt}\left(\underline{T}_{i}{}^{0}-N_{ik}{}^{l}\underline{\Delta}^{k}{}_{l}{}^{0}\right)  &=&R_{ijk}{}^{l}\underline{\Delta}^{k}{}_{l}{}^{j}+{\stackrel{\{\,\}}{\Gamma}}_{ik}{}^{l}\underline{T}_{l}{}^{k}+{\stackrel{\{\,\}}{\Gamma}}_{ik}{}^{l}{}_{,a} \, \underline{T}^{a}{}_{l}{}^{k}-N_{ik}{}^{l}\left( \Gamma _{jl}{}^{m}\underline{\Delta}^{k}{}_{m}{}^{j}\right.   \nonumber \\
&&\left. -\Gamma _{mj}{}^{k}\underline{\Delta}^{j}{}_{l}{}^{m}\right) -N_{ik}{}^{l}\,_{,a}\left(\underline{T}^{a}{}_{l}{}^{k}-\underline{t}^{ak}{}_{l}+v^{a}\underline{\Delta}^{k}{}_{l}{}^{0}\right)   \label{DY_deriv_inter_3} \\
&=&{\stackrel{\{\,\}}{\Gamma}}_{ik}{}^{l}\underline{T}_{l}{}^{k}+{\stackrel{\{\,\}}{\Gamma}}_{ik}{}^{l}{}_{,a} \, \underline{T}^{a}{}_{l}{}^{k}+\underline{\Delta}^{k}{}_{l}{}^{j}\left( R_{ijk}{}^{l}-\Gamma _{jp}{}^{l}N_{ik}{}^{p}+\Gamma_{jk}{}^{p}N_{ip}{}^{l}\right.   \nonumber \\
&&\left. -N_{ik}{}^{l}{}_{,j}+{\stackrel{\{\,\}}{\Gamma}}_{ji}{}^{p}N_{pk}{}^{l}-{\stackrel{\{\,\}}{\Gamma}}_{ji}{}^{p}N_{pk}\,^{l}\right) .  \label{DY_deriv_inter_4}
\end{eqnarray}
In the last step we made use of (\ref{DY_pd_prop_eq_5}) in order to replace the terms in the last brace in the second line of (\ref{DY_deriv_inter_3}). Furthermore, we added a ``0'' dummy term, i.e., the last two terms in the second line of (\ref{DY_deriv_inter_4}). We proceed by replacing the curvature by its decomposition, i.e.,
\begin{equation}
R_{ijk}{}^{l}={\stackrel{\{\,\}}{R}}_{ijk}{}^{l}+{\stackrel{\{\,\}}{\nabla}}_{j}N_{ik}{}^{l}-{\stackrel{\{\,\}}{\nabla}}_{i}N_{jk}{}^{l}+N_{ip}{}^{l}N_{jk}{}^{p}-N_{jp}{}^{l}N_{ik}{}^{p}, \label{curvature_decomposition}
\end{equation}
equation (\ref{DY_deriv_inter_4}) then turns into
\begin{equation}
\frac{d}{dt}\left(\underline{T}_{i}{}^{0}-N_{ik}{}^{l}\underline{\Delta}^{k}{}_{l}{}^{0}\right) ={\stackrel{\{\,\}}{\Gamma}}_{ik}{}^{l}\underline{T}_{l}{}^{k}+{\stackrel{\{\,\}}{\Gamma}}_{ik}{}^{l}{}_{,j} \, \underline{T}^{j}{}_{l}{}^{k}+\underline{\Delta}^{k}{}_{l}{}^{j}\left({\stackrel{\{\,\}}{R}}_{ijk}{}^{l}-{\stackrel{\{\,\}}{\Gamma}}_{ji}{}^{p}N_{pk}{}^{l}-{\stackrel{\{\,\}}{\nabla}}_{i}N_{jk}{}^{l}\right).  \label{DY_deriv_inter_5}
\end{equation}
We rewrite (\ref{DY_pd_prop_eq_2}) with the help of (\ref{DY_pd_prop_eq_5}):
\begin{equation}
\underline{T}_{l}{}^{k}=\frac{d}{dt}\underline{T}^{k}{}_{l}{}^{0}+v^{k}\underline{T}_{l}{}^{0}-{\stackrel{\{\,\}}{\Gamma}}_{lp}{}^{m}\underline{T}^{k}{}_{m}{}^{p}+N_{lp}{}^{m}\left(\underline{\Delta}^{p}{}_{m}{}^{k}-v^{k}\underline{\Delta}^{p}{}_{m}{}^{0}\right) .  \label{DY_deriv_inter_6}
\end{equation}
Contracting this equation with the Levi-Civita connection and introducing another ``0'' dummy term yields
\begin{eqnarray}
{\stackrel{\{\,\}}{\Gamma}}_{ik}{}^{l}\underline{T}_{l}{}^{k} &=&\frac{d}{dt}\left({\stackrel{\{\,\}}{\Gamma}}_{ik}{}^{l}\underline{T}^{k}{}_{l}{}^{0}\right) +{\stackrel{\{\,\}}{\Gamma}}_{ik}{}^{l}v^{k}\left(\underline{T}_{l}{}^{0}-N_{lp}{}^{m}\underline{\Delta}^{p}{}_{m}{}^{0}-{\stackrel{\{\,\}}{\Gamma}}_{lp}{}^{m}\underline{T}^{p}{}_{m}{}^{0}\right)   \nonumber \\
&&-v^{a}{\stackrel{\{\,\}}{\Gamma}}_{ik}{}^{l}{}_{,a}\underline{T}^{k}{}_{l}{}^{0}-{\stackrel{\{\,\}}{\Gamma}}_{ik}{}^{l}{\stackrel{\{\,\}}{\Gamma}}_{lp}{}^{m}\stackrel {(c)}{\underline{T}}{}^{k}{}_{m}{}^{p}+{\stackrel{\{\,\}}{\Gamma}}_{ik}{}^{l}N_{lp}{}^{m}\underline{\Delta}^{p}{}_{m}{}^{k}
\nonumber\\ &&+ \left({\stackrel{\{\,\}}{\Gamma}}_{ik}{}^{l}{\stackrel{\{\,\}}{\Gamma}}_{lp}{}^{m} - {\stackrel{\{\,\}}{\Gamma}}_{ip}{}^{l}{\stackrel{\{\,\}}{\Gamma}}_{lk}{}^{m}\right)\underline{T}^p{}_{m}{}^{0}\,v^k.  \label{DY_deriv_inter_7}
\end{eqnarray}
With this result at hand we can replace the first term on the rhs of (\ref{DY_deriv_inter_5}), i.e.,
\begin{eqnarray}
&&\frac{d}{dt}\left(\underline{T}_{i}{}^{0}-N_{ik}{}^{l}\underline{\Delta}^{k}{}_{l}{}^{0}-{\stackrel{\{\,\}}{\Gamma}}_{ik}{}^{l}\underline{T}^{k}{}_{l}{}^{0}\right)-
{\stackrel{\{\,\}}{\Gamma}}_{ik}{}^{l}v^{k}\left(\underline{T}_{l}{}^{0}-N_{lp}{}^{m}\underline{\Delta}^{p}{}_{m}{}^{0}-{\stackrel{\{\,\}}{\Gamma}}_{lp}{}^{m}\underline{T}^{p}{}_{m}{}^{0}\right)   \nonumber \\
&=&\left({\stackrel{\{\,\}}{\Gamma}}_{ik}{}^{l}{}_{,j}-{\stackrel{\{\,\}}{\Gamma}}_{ij}{}^{p}{\stackrel{\{\,\}}{\Gamma}}_{pk}{}^{l}\right)\stackrel{(c)}{\underline{T}}{}^{j}{}_{l}{}^{k}+\underline{\Delta}^{k}{}_{l}{}^{j}\left({\stackrel{\{\,\}}{R}}_{ijk}{}^{l}-{\stackrel{\{\,\}}{\nabla}}_{i}N_{jk}{}^{l}\right)+{\stackrel{\{\,\}}{R}}_{kji}{}^{l}\,\underline{T}^k{}_{l}{}^{0}\,v^j.  \label{DY_deriv_inter_8}
\end{eqnarray}
If we introduce a new quantity
\begin{equation}
{\cal P}_{i}:=\underline{T}_{i}{}^{0}-N_{ik}{}^{l}\underline{\Delta}^{k}{}_{l}{}^{0}-{\stackrel{\{\,\}}{\Gamma}}_{ik}{}^{l}\underline{T}^{k}{}_{l}{}^{0}  \label{Ptot}
\end{equation}
as a generalized total 4-momentum, equation (\ref{DY_deriv_inter_8}) can be written in a more compact form as follows:
\begin{equation}
{\stackrel{\{\,\}}{\nabla}}_{v}{\cal P}_{i}={\stackrel{\{\,\}}{R}}_{kji}{}^{l}\,\underline{T}^k{}_{l}{}^{0}\,v^j +
\left({\stackrel{\{\,\}}{\Gamma}}_{ik}{}^{l}{}_{,j}-{\stackrel{\{\,\}}{\Gamma}}_{ij}{}^{p}{\stackrel{\{\,\}}{\Gamma}}_{pk}{}^{l}\right)\stackrel{(c)}{\underline{T}}{}^{j}{}_{l}{}^{k}+\underline{\Delta}^{k}{}_{l}{}^{j}\left({\stackrel{\{\,\}}{R}}_{ijk}{}^{l}-{\stackrel{\{\,\}}{\nabla}}_{i}N_{jk}{}^{l}\right).  \label{DY_deriv_inter_9}
\end{equation}
By using the Ricci identity ${\stackrel{\{\,\}}{R}}_{jki}{}^{l}+{\stackrel{\{\,\}}{R}}_{kij}{}^{l}+{\stackrel{\{\,\}}{R}}_{ijk}{}^{l}=0$ and the fact that the convective part of first integrated moment of the canonical-momentum is antisymmetric in the upper two indices, i.e., $\stackrel {(c)}{\underline{T}}{}^{k}{}_{m}{}^{p}=\stackrel {(c)}{\underline{T}}{}^{[k}{}_{m}{}^{p]}$, we can recast (\ref{DY_deriv_inter_9}) into
\begin{equation}
{\stackrel{\{\,\}}{\nabla}}_{v}{\cal P}_{i}={\stackrel{\{\,\}}{R}}_{kji}{}^{l}\,\underline{T}^k{}_{l}{}^{0}\,v^j + 
\left({\stackrel{\{\,\}}{R}}_{ijk}{}^{l}-{\stackrel{\{\,\}}{\nabla}}_{i}N_{jk}{}^{l}\right)\underline{\Delta}^{k}{}_{l}{}^{j}+{\stackrel{\{\,\}}{R}}_{ijk}{}^{l}\stackrel {(c)}{\underline{T}}{}^{k}{}_{l}{}^{j}.  \label{DY_pd_prop_eq_1a}
\end{equation}
This equation represents the rewritten form of (\ref{DY_pd_prop_eq_1}) and should be compared to (\ref{final_res_analogous_ys80}) in the previous section.

It is worthwhile to notice the general feature that characterizes the coupling between the physical objects (currents) with the geometrical objects (metric, connection, and the derived quantities). Namely, the {\it intrinsic} current (the one that is truly {\it microscopic}, which arises from the averaging over the medium with the elements with microstructure, i.e., that possess internal degrees of freedom) couples to the {\it post-Riemannian} geometric quantities, see the second term on the rhs of (\ref{Ptot}) and the first term on the rhs of (\ref{DY_pd_prop_eq_1a}). In contrast to this, the {\it orbital} canonical energy-momentum (which is induced by the {\it macroscopic} dynamics of the rotating and deformable body) is only coupled to the purely Riemannian geometric variables and never couples to the post-Riemannian geometry, see the last terms on the right-hand sides of (\ref{Ptot}) and (\ref{DY_pd_prop_eq_1a}). This observation represents a generalization of the result of Yasskin and Stoeger \cite{Stoeger:Yasskin:1980}, in other words, it proves that the possible presence of the post-Riemannian geometry (in particular, of the torsion and the nonmetricity) can only be tested with the help of the bodies that are constructed from media with microstructure (spin, dilaton charge, and intrinsic shear). Test particles composed from usual matter, i.e., without microstructure, are not affected by the post-Riemannian geometry, and they thus cannot be used for the detection of the torsion and the nonmetricity. 

In order to get a better understanding of this fact, we will consider several special cases of the metric-affine geometry in the subsequent sections, moving from a general non-Riemannian geometry back to the Riemannian one.

\section{Relation between the integrated moments}\label{relation_momenta_sec}

In different situations, it is technically convenient to use different definitions of the integrated moments (see also \cite{Chen:1993} for the behaviour under infinitesimal coordinate transformations). However, directly from the definitions (\ref{int_moments_definitions}) and (\ref{DY_int_moments_definitions}) we can establish relations between two sets of the moments.

Starting with the identity $\tilde{t}^{ij} = g^{jk}\,\tilde{t}^i{}_k$, we expand the metric in the same way as the other geometric quantities (\ref{DY_geom_quant_taylor}),
\begin{equation}
\left. g^{jk}\right|_{x} = \left. g^{jk}\right|_{Y} + \left. \delta x^a \,g^{jk}{}_{,a}\right|_{Y}+\cdots ,
\end{equation}
and then by integration over the world tube, in the {\it pole-dipole approximation} we find
\begin{equation}
\overline{t}^{ij} = \underline{t}^{ij} - 2\,{\stackrel{\{\,\}}{\Gamma}}_l{}^{(kj)} \,\underline{t}^{li}{}_k.\label{tt}
\end{equation}
We used here the metricity condition $g^{jk}{}_{,l} = - {\stackrel{\{\,\}}{\Gamma}}_{ln}{}^j\,g^{nk} - {\stackrel{\{\,\}}{\Gamma}}_{ln}{}^k\,g^{jn}$.

Analogously, we have for the integrated canonical energy-momentum
\begin{equation}
\overline{T}^{ij} = \underline{T}^{ij} - 2\,{\stackrel{\{\,\}}{\Gamma}}_l{}^{(ik)} \,\underline{T}^{l}{}_k{}^j.\label{TT}
\end{equation}
The ``inverse" formulas read
\begin{eqnarray}
\underline{t}^i{}_j &=& \overline{t}^i{}_j + 2\,{\stackrel{\{\,\}}{\Gamma}}{}_{l(jk)}\,\overline{t}^{lik},\label{tt1}\\
\underline{T}_i{}^j &=& \overline{T}_i{}^j + 2\,{\stackrel{\{\,\}}{\Gamma}}{}_{l(ik)}\,\overline{T}^{lkj}. \label{TT1}
\end{eqnarray}

Hence, in the pole-dipole approximation, the integrated hypermomenta and the first moments of the canonical and metrical energy-momenta in both sets are the same:
\begin{eqnarray}
\underline{\Delta}^{ijk} &=& \overline{\Delta}^{ijk},\label{Dud}\\
\underline{t}^{ijk} &=& \overline{t}^{ijk},\label{tud}\\
\underline{T}^{ijk} &=& \overline{T}^{ijk}.\label{Tud}
\end{eqnarray}
With the help of (\ref{tud}) and (\ref{Tud}), we can verify the consistency of the relations (\ref{tt}) and (\ref{tt1}), as well as (\ref{TT}) and (\ref{TT1}).

For single-pole test particles, the corresponding integrated energy-momenta coincide since the last terms in (\ref{tt})-(\ref{TT1}) vanish.

\section{Conclusions \& Outlook}\label{conclusions_outlook_sec}

In this work we derived the equations of motion for test particles in metric-affine gravity from the conservation laws of the theory with the help of a multipole formalism. Apart from the general form of the equations of motion, we explicitly presented the propagation equations for pole-dipole test particles. Our results are valid for a very large class of gravitational theories, i.e., all theories which fit into the framework of metric-affine gravity. The equations derived in this work should be used to systematically study the motion of test particles with spin, shear, dilation, and rotation within alternative gravitational theories in a non-Riemannian context. Our results generalize previous analyses \cite{Hehl:1971,Trautman:1972,Stoeger:Yasskin:1979,Stoeger:Yasskin:1980,Hehl:Neeman:1997}, which were carried out in the context of general relativity, Einstein-Cartan theory, and within Poincar\'{e} gauge theory.  

\subsection{Special cases}

In this section we discuss several special cases within our framework by either making assumptions about the internal structure of the test particles, or by constraining the background geometry. The full agreement, in some special cases, with the well-known results from general relativity and Poincar\'{e} gauge theory demonstrates the consistency of our framework. 

\subsubsection{Equations for a single-pole particle in metric-affine gravity}

Let us consider the propagation equations for a single-pole test particle in metric-affine gravity, i.e., the set (\ref{pd_prop_eq_1})-(\ref{pd_prop_eq_5}) with vanishing dipole contributions:
\begin{eqnarray}
\frac{d}{dt}\overline{T}^{i0} &=&R^{i}\,_{jkl}\overline{\Delta}^{klj}-N^{i}{}_{kl}\overline{T}^{lk}-{\stackrel{\{\,\}}{\Gamma}}_{kj}{}^{i}\overline{T}^{(kj)}+N^{i}{}_{kl}\overline{t}^{kl},  \label{mag_single_pole_1} \\
v^{a}\overline{T}^{i0} &=&\overline{T}^{ia},  \label{mag_single_pole_2} \\
\frac{d}{dt}\overline{\Delta}^{kl0} &=&N_{mj}{}^{k}\overline{\Delta}^{jlm}-{\stackrel{\{\,\}}{\Gamma}}_{mj}{}^{k}\overline{\Delta}^{mlj}-{\stackrel{\{\,\}}{\Gamma}}_{mj}{}^{l}\overline{\Delta}^{k(mj)}-N^{jlm}\overline{\Delta}^{k}{}_{mj}-\overline{T}^{lk}+\overline{t}^{kl},  \label{mag_single_pole_3} \\
v^{a}\overline{\Delta}^{kl0} &=&\overline{\Delta}^{kla}.  \label{mag_single_pole_4}
\end{eqnarray}
It is a common folklore that in generalized gravity theories the equation of motion for single-pole test particles is given by some kind of ``generalized'' geodesic equation. By generalized we mean an equation which has the same form as the geodesic equation, i.e., the equation of motion for single-pole test particles in general relativity, but in which the Levi-Civita connection has been replaced by the full (non-Riemannian) connection. The result in (\ref{mag_single_pole_1})-(\ref{mag_single_pole_4}) clearly demonstrates that such an assumption is \textit{not} substantiated.

\paragraph{Particles without intrinsic hypermomentum}

If we perform a further specialization by considering only test particles without intrinsic hypermomentum, the set (\ref{mag_single_pole_1})-(\ref{mag_single_pole_4}) turns into 
\begin{eqnarray}
\frac{d}{dt}\overline{T}^{i0} &=&-N^{i}{}_{kl}\overline{T}^{lk}-{\stackrel{\{\,\}}{\Gamma}}_{kj}{}^{i}\overline{T}^{(kj)}+N^{i}{}_{kl}\overline{t}^{kl}, \label{mag_single_pole_no_hyp_1} \\
v^{a}\overline{T}^{i0} &=&\overline{T}^{ia},  \label{mag_single_pole_no_hyp_2} \\
\overline{T}^{lk} &=&\overline{t}^{kl}.  \label{mag_single_pole_no_hyp_3}
\end{eqnarray}
Of course the first and the last term on the rhs of (\ref{mag_single_pole_no_hyp_1}) cancel because of (\ref{mag_single_pole_no_hyp_3}) and the equation of motion for a test particle without intrinsic hypermomentum is then given by the regular geodesic equation [in the next section we explicitly show how one can recover the geodesic equation from the set (\ref{mag_single_pole_no_hyp_1})-(\ref{mag_single_pole_no_hyp_3})]. This generalizes the well-known result from Poincar\'{e} gauge theory to metric-affine gravity, i.e., a test particle without intrinsic hypermomentum will \textit{not} ``feel'' the torsion or the nonmetricity of the underlying spacetime. Hence, test particles without intrinsic spin, shear, or dilation current are not suitable for mapping the non-Riemannian features of spacetime. Accordingly, current experiments like Gravity Probe-B \cite{GPBweb} are \textit{not} suitable for the detection of torsion in contrast to what is sometimes claimed by other authors. At this point, one should mention that a coupling between torsion and matter without intrinsic spin currents may be achieved in some nonstandard gravity theory, although the authors of the present paper are not aware of any viable candidate for such a theory. For any theory which fits into the very general and well-motivated framework of metric-affine gravity, e.g., Poincar\'{e} gauge theory and Einstein-Cartan theory, such a coupling will \textit{not} occur.

\subsubsection{Recovering the geodesic equation}

In this section we explicitly show that the single-pole equations of motion for a test particle {\it without} intrinsic hypermomentum take the form of the usual geodesic equation. The set (\ref{mag_single_pole_no_hyp_1})-(\ref{mag_single_pole_no_hyp_3}) reduces to
\begin{eqnarray}
\frac{d}{dt}\overline{T}^{i0} &=&-{\stackrel{\{\,\}}{\Gamma}}_{kj}{}^{i}\overline{T}^{(kj)},  \label{geo1} \\
\overline{T}^{ia} &=&v^{a}\overline{T}^{i0},  \label{geo2} \\
\overline{T}^{lk} &=&\overline{t}^{kl}.  \label{geo3}
\end{eqnarray}
Now lets us introduce the velocity $u^{a}:=dY^{a}/ds$ along the world line of the particle. Note that $u^{0}=dt/ds,$ $ds^{2}=g_{ab}dY^{a}dY^{b},$ and remember that $Y^{a}\left(t\right)=x^{a}\left( Y(t)\right) =t\delta _{0}^{a},$ $d/dt=v^{a}\partial _{a},$ $u^{a}u_{a}=1,$ $v^{a}=dY^{a}/dt$. With this definition we can rewrite (\ref{geo1}) and (\ref{geo2}) as follows: 
\begin{eqnarray}
\frac{d}{ds}\overline{T}^{i0}+{\stackrel{\{\,\}}{\Gamma}}_{kj}{}^{i}u^{0}\overline{T}^{kj} &=&0,  \label{geo1_2} \\
u^{0}\overline{T}^{ia} &=&u^{a}\overline{T}^{i0}.  \label{geo2_2}
\end{eqnarray}
Setting $i=0$ in the last equation and reinsertion into (\ref{geo1_2}), together with the definition $m:=\overline{T}^{00}/\left( u^{0}\right)^{2}$, yields $\overline{T}^{ia}=mu^{i}u^{a}$. This in turn can be used to rewrite (\ref{geo1_2}) as follows:
\begin{equation}
\frac{d}{ds}\left( mu^{i}\right) +{\stackrel{\{\,\}}{\Gamma}}_{kj}{}^{i}mu^{k}u^{j}=0.  \label{geo1_3}
\end{equation}
Multiplication of this equation by $u_{i}$ and remembering that $u^{b}{\stackrel{\{\,\}}{\nabla}}_b u^{a}=\left( u^{a}{}_{,b}+{\stackrel{\{\,\}}{\Gamma}}_{cb}{}^{a}u^{c}\right) u^{b}$, $du^{a}/ds=u^{a}{}_{,b}u^{b},$ $u_{a} {\stackrel{\{\,\}}{\nabla}}_b u^{a}=0$ yields
\begin{equation}
\frac{dm}{ds}u^{i}u_{i}+m u_{k} u^{j} {\stackrel{\{\,\}}{\nabla}}_ju^{k}=0\quad \Longrightarrow \quad 
\frac{dm}{ds}=0.  \label{sp_eom_1}
\end{equation}
When we use this result in (\ref{geo1_3}) we end up with
\begin{equation}
\frac{du^{i}}{ds}+{\stackrel{\{\,\}}{\Gamma}}_{kj}{}^{i}u^{k}u^{j}=0,  \label{sp_eom_2}
\end{equation}
which is the geodesic equations. Hence, {\it in metric-affine gravity single-pole test particles without intrinsic hypermomentum, i.e., without spin, shear, and dilation currents, move in exactly the same way as test particles in general relativity}. We stress that {\it no} constraining assumptions about the geometry of the background spacetime have been made in order to derive this result. Equation (\ref{sp_eom_2}) is valid in a completely general metric-affine spacetime, i.e., the background can be a non-Riemannian one with nonvanishing torsion and nonmetricity, the test particle just does {\it not} feel this geometric features as long as it does not posses any ``microstructure'' in the form of a nonvanishing intrinsic hypermomentum.

In later sections we are also going to discuss the equations of motion for some special cases in which we impose an a priori restriction on the geometry of the background spacetime.

\subsubsection{Recovering the Mathisson-Papapetrou equations}

Also the well-known propagation equations for a classical pole-dipole test particle can be easily recovered in our framework. For particles without intrinsic hypermomentum in a Riemannian background the propagation equations in (\ref{pd_prop_eq_1})-(\ref{pd_prop_eq_5}) turn into
\begin{eqnarray}
\frac{d}{dt}\overline{T}^{i0} &=&-{\stackrel{\{\,\}}{\Gamma}}_{kj}{}^{i}\overline{T}^{(kj)}-{\stackrel{\{\,\}}{\Gamma}}_{kj}{}^{i}{}_{,a}\overline{T}^{a(kj)}, \label{matpap1} \\
\frac{d}{dt}\overline{T}^{ai0} &=&\overline{T}^{ia}-v^{a}\overline{T}^{i0}-{\stackrel{\{\,\}}{\Gamma}}_{kj}{}^{i}\overline{T}^{a(kj)},  \label{matpap2} \\
v^{a}\overline{T}^{bi0}+v^{b}\overline{T}^{ai0} &=&\overline{T}^{bia}+\overline{T}^{aib},  \label{matpap3} \\
\overline{T}^{lk} &=&\overline{t}^{kl},  \label{matpap4} \\
\overline{T}^{alk} &=&\overline{t}^{akl}.  \label{matpap5}
\end{eqnarray}
These equations are exactly the equations of motion for a pole-dipole particle described by Papapetrou in (3.2)-(3.4) of \cite{Papapetrou:1951:3}. This result clearly demonstrates the consistency and generality of our framework.

\subsubsection{Propagation equations in a Weyl-Cartan spacetime}

The Weyl-Cartan spacetime is characterized by a special type of nonmetricity, namely, when the 1-form of the nonmetricity $Q_{\alpha\beta}= g_{\alpha\beta}\,Q$ reduces to just the Weyl covector $Q = Q_idx^i$. Correspondingly, the distorsion 1-form then reduces to 
\begin{equation}
N_\alpha{}^\beta = -\,{\frac 12}\,\delta_\alpha^\beta\,Q + K_\alpha{}^\beta,
\label{NQK}
\end{equation}
where the contortion $K_{\alpha\beta} = -\,K_{\beta\alpha} := N_{[\alpha\beta]}$ is just the antisymmetric piece of the distorsion (note, however, that $K_\alpha{}^\beta$ is constructed from both the torsion and the Weyl nonmetricity). In components, we have explicitly $N_{i\alpha}{}^\beta = -\,{\frac{1}{2}}\,\delta_\alpha^\beta\,Q_i + K_{i\alpha}{}^\beta$. 

Using relation (\ref{NQK}), we derive the propagation equations for test particles on the background of the Weyl-Cartan spacetime:
\begin{eqnarray}
{\stackrel {\{\,\}} {\nabla}}{}_v\,{\cal P}_i &=& {\stackrel{\{\,\}}{R}}_{kji}{}^{l}\,\underline{T}^k{}_{l}{}^{0}\,v^j + \left(\stackrel{\{\,\}}{R}_{ijk}{}^l - {\stackrel {\{\,\}}{\nabla}}{}_i\,K_{jk}{}^{l}\right) \underline{\tau}{}^k{}_l{}^j+ \stackrel{\{\,\}}{R}_{ijk}{}^l\, {\stackrel {(c)}{\underline{T}}}{}^k{}_l{}^j + {\frac 12}({\stackrel {\{\,\}} {\nabla}}{}_iQ_j)\,\underline{Z}^j,\label{DY_pd_prop_eq_1b}\\
\underline{T}_k{}^i &=& v^i\,\underline{P}_k + {\frac {d}{dt}}\,\underline{L}^i{}_k - {\stackrel{\{\,\}}{\Gamma}}_{kj}{}^{l} \,\underline{T}^i{}_l{}^j + K_{kj}{}^{l}\,{\stackrel {(c)}{\underline{\tau}}}{}^j{}_l{}^i - {\frac 12}\,Q_k\,{\stackrel {(c)}{\underline{Z}}}{}^i,\label{DY_pd_prop_eq_2b}\\
{\stackrel {(c)}{\underline{T}}}{}^{(a}{}_i{}^{b)} &=& 0,\label{DY_pd_prop_eq_3b}\\
{\stackrel {\{\,\}} {\nabla}}_v\,\underline{Y}^i{}_k &=& -\,\underline{T}_k{}^i+ \underline{t}^i{}_k - {\stackrel {\{\,\}} \Gamma}_{jl}{}^{i}\,{\stackrel{(c)}{\underline{\Delta}}}{}^l{}_k{}^j + {\stackrel {\{\,\}} \Gamma}_{jk}{}^{l}\,{\stackrel {(c)}{\underline{\Delta}}}{}^i{}_l{}^j +K_{jl}{}^{i}\,\underline{\Delta}^l{}_k{}^j - K_{jk}{}^{l}\,\underline{\Delta}{}^i{}_l{}^j,\label{DY_pd_prop_eq_4b}\\
{\stackrel {(c)}{\underline{\Delta}}}{}^k{}_l{}^a &=& \underline{T}^a{}_l{}^k - \underline{t}^{ak}{}_l.\label{DY_pd_prop_eq_5b}
\end{eqnarray}
Here ${\cal P}_i = \underline{P}_i + {\frac 12}\,Q_i\,\underline{Z} - K_{ik}{}^{l}\,\underline{\tau}^k{}_l - {\stackrel{\{\,\}}{\Gamma}}_{ik}{}^{l}\,\underline{L}^k{}_l$. 

\paragraph{Single-pole particles} 

For the single-pole case (when all of the first integrated moments vanish), we find a surprisingly simple system
\begin{eqnarray}
{\stackrel {\{\,\}} {\nabla}}{}_v\,\underline{P}_i + K_{ij}{}^k\,v^j\underline{P}_k &=& R_{ijk}{}^l\,v^j\underline{\tau}^k{}_l + {\frac 12}f_{ij}v^j\underline{Z} - {\frac 12}\,Q_i\,{\frac {d\underline{Z}}{dt}},\label{WCsingle1}\\
\underline{T}_k{}^i &=& v^i\,\underline{P}_k,\label{WCsingle2}\\
\nabla_v\,\underline{Y}^i{}_k &=& -\,\underline{T}_k{}^i + \underline{t}^i{}_k,\label{WCsingle3}\\
{\stackrel {(c)}{\underline{\Delta}}}{}^k{}_l{}^a &=& 0.\label{WCsingle4}
\end{eqnarray}
Here we introduced $f_{ij} := \partial_iQ_j - \partial_jQ_i$. Thus, provided a test particle has a nontrivial integrated dilaton charge $\underline{Z}$, it will be affected in the Weyl-Cartan spacetime by the Lorentz--type force represented by the second term on the rhs of the propagation equation (\ref{WCsingle1}). If, in addition, the test particle has a nontrivial spin $\underline{\tau}^k{}_l$, the latter will be affected by the Mathisson-Papapatrou--type force which is determined by the Weyl-Cartan curvature, as described by the first term on the rhs (\ref{WCsingle1}). 

\subsubsection{Propagation equations in a Weyl spacetime}

Weyl \cite{Weyl:1918,Weyl:1919,Weyl:1923} was the first who noticed a similarity between the electromagnetic vector potential and the nonmetricity covector $Q_i$. Indeed, this is also manifested in the equations of motion, as becomes clear from the rhs of equation (\ref{WCsingle1}). However, an essential difference is that the Weyl nonmetricity may interact with the dilaton charge and not with the electromagnetic charge. 

The Weyl geometry arises as a special case of the Weyl-Cartan spacetime, when the torsion $S_{ij}{}^k := \Gamma_{ij}{}^k - \Gamma_{ji}{}^k=0$ is equal zero.\footnote{Our notation for the spacetime torsion is different from \cite{Hehl:1995}. Since we reserved the symbol $T$ for energy-momentum related objects, the torsion tensor is here denoted by the symbol $S$ as in the old review \cite{Hehl:1976}.} In this case the distorsion is still given by (\ref{NQK}), but the contortion is expressed in terms of the Weyl covector only:
\begin{equation}
K_{ij}{}^k = {\frac 12}\left(g_{ij}\,Q^k - \delta_i^k\,Q_j\right).\label{KW}
\end{equation}
The propagation equations in the Weyl spacetime are formally the same as (\ref{DY_pd_prop_eq_1b})-(\ref{DY_pd_prop_eq_5b}) where we have to substitute the contortion (\ref{KW}). Analogously, the dynamics of single-pole test particles is described in the Weyl spacetime by (\ref{WCsingle1})-(\ref{WCsingle3}) with (\ref{KW}) inserted. 

\subsubsection{Propagation equations in a Riemann-Cartan spacetime}

The Riemann-Cartan spacetime arises from the Weyl-Cartan geometry for the case of vanishing nonmetricity, $Q_i = 0$. The distorsion then coincides with the contortion and is constructed only from the torsion: $N_{ijk} = K_{ijk} = {\frac 12}\left(S_{jki} + S_{ikj} + S_{jik}\right)$. 

The propagation equations for pole-dipole particles in Riemann-Cartan spacetime are easily derived by putting $Q_i = 0$ in equations (\ref{DY_pd_prop_eq_1b})-(\ref{DY_pd_prop_eq_5b}). We will not write these equations explicitly. 

\paragraph{Single-pole particles}

In order to discuss the propagation equations for single-pole particles, we again introduce the 4-velocity $u^{a}:=dY^{a}/ds$ along the world line of the particle. With $u^{0}=dt/ds$ and $ds^{2}=g_{ab}dY^{a}dY^{b},$ we have $u_au^a = 1$ (note that $u^a = u^0v^a$). Then, it is straightforward to verify that in the Riemann-Cartan spacetime equations (\ref{WCsingle1})-(\ref{WCsingle3}) reduce to
\begin{eqnarray}
\dot{\underline{P}}{}_i &=& S_{ij}{}^k\,u^j\underline{P}_k + R_{ijk}{}^l\,u^j\underline{\tau}^k{}_l,\label{WCsingle1a}\\
u^0\underline{T}_k{}^i &=& u^i\,\underline{P}_k,\label{WCsingle2a}\\
\dot{\underline{\tau}}{}_{ij} &=& u_{[i}\,\underline{P}_{j]},\label{WCsingle3a}\\
\dot{\underline{Y}}{}_{(ij)} &=& u^0\left(\underline{t}_{(ij)} -\,\underline{T}_{(ij)}\right).\label{WCsingle4a}
\end{eqnarray}
Here we denoted the covariant (Riemann-Cartan) derivative along the trajectory by a dot: ``$\, \dot{} \,$''$= D/ds = u^i\nabla_i$. 

It is satisfactory to see that with (\ref{WCsingle1a}) and (\ref{WCsingle3a}) we recover the usual equations of motion for a test particle with mass and spin in the Riemann-Cartan spacetime \cite{Hehl:1971,Trautman:1972,Hehl:1976}. One should note, however, that we are still in the framework of the metric-affine gravity in which a test particle carries, besides the mass and spin, also the dilaton charge and the intrinsic shear. The latter integrated characteristics are described by the symmetric part of the intrinsic hypermomentum $\underline{Y}{}_{(ij)}$. The dynamics of these quantities is determined by equation (\ref{WCsingle4a}) which is completely decoupled from the other propagation equations. In other words, they do not affect the motion of a test particle in the Riemann-Cartan spacetime, and the trajectory is completely defined by the behavior of the integrated 4-momentum $\underline{P}_i$ and the integrated spin $\underline{\tau}^k{}_l$. 

Let us contract equation (\ref{WCsingle3a}) with $u^i$. This then yields the explicit form of the integrated 4-momentum, 
\begin{equation}
\underline{P}_j = m\,u_j + 2u^i\,\dot{\underline{\tau}}{}_{ij},\label{Putau}
\end{equation}
where we introduced the notation for the rest mass of the body $m := u^i \underline{P}_i$ (i.e., the momentum projected to the rest frame). By substituting this back into (\ref{WCsingle3a}) we obtain the dynamical equation for the spin
\begin{equation}
\dot{\underline{\tau}}{}_{ij} - u_iu^k\dot{\underline{\tau}}{}_{kj}
+ u_ju^k\dot{\underline{\tau}}{}_{ki} = 0.\label{dottau} 
\end{equation}

\subsubsection{Propagation equations in a Riemannian spacetime}

When all the post-Riemannian geometric objects are trivial (no torsion and no nonmetricity, i.e., $N_\alpha{}^\beta = 0$), the propagation equations on the purely Riemannian spacetime reduce to
\begin{eqnarray}
{\stackrel {\{\,\}} {\nabla}}{}_v\,{\cal P}_i &=& {\stackrel{\{\,\}}{R}}_{kji}{}^{l}\,\underline{T}^k{}_{l}{}^{0}\,v^j +
\stackrel{\{\,\}}{R}_{ijk}{}^l\Big(\underline{\tau}{}^k{}_l{}^j + {\stackrel{(c)}{\underline{T}}}{}^k{}_l{}^j\Big),\label{DY_pd_prop_eq_1c}\\
\underline{T}_k{}^i &=& v^i\,\underline{P}_k + {\frac {d}{dt}}\,\underline{L}^i{}_k - {\stackrel{\{\,\}}{\Gamma}}_{kj}{}^{l} \,\underline{T}^i{}_l{}^j,\label{DY_pd_prop_eq_2c}\\
{\stackrel {(c)}{\underline{T}}}{}^{(a}{}_i{}^{b)} &=& 0, \label{DY_pd_prop_eq_3c}\\
{\stackrel {\{\,\}} {\nabla}}_v\,\underline{Y}^i{}_k &=& -\,\underline{T}_k{}^i + \underline{t}^i{}_k - {\stackrel {\{\,\}} \Gamma}_{jl}{}^{i} \,{\stackrel{(c)}{\underline{\Delta}}}{}^l{}_k{}^j + {\stackrel {\{\,\}} \Gamma}_{jk}{}^{l}\,{\stackrel {(c)}{\underline{\Delta}}}{}^i{}_l{}^j,\label{DY_pd_prop_eq_4c}\\
{\stackrel {(c)}{\underline{\Delta}}}{}^k{}_l{}^a &=& \underline{T}^a{}_l{}^k - \underline{t}^{ak}{}_l.\label{DY_pd_prop_eq_5c}
\end{eqnarray}
Here ${\cal P}_i = \underline{P}_i - {\stackrel{\{\,\}}{\Gamma}}_{ik}{}^{l} \,\underline{L}^k{}_l$. 

\paragraph{Single-pole particles}

For the single-pole particles with vanishing intrinsic hypermomentum this simplifies to 
\begin{eqnarray}
{\stackrel {\{\,\}} {\nabla}}{}_v\,\underline{P}_i &=& 0,\label{DY_pd_prop_eq_1d}\\
\underline{T}_k{}^i &=& v^i\,\underline{P}_k,\label{DY_pd_prop_eq_2d}\\
\underline{T}_k{}^i &=& \underline{t}^{i}{}_k.\label{DY_pd_prop_eq_4d}
\end{eqnarray}
The resulting trajectories are geodesics. 

\subsubsection{Propagation equations in a Riemannian spacetime (alternative form)}

For completeness let us also determine the explicit form of the propagation equations using the upper-index convention for the integrated moments, for the special case of a Riemannian background. From (\ref{pd_prop_eq_1})-(\ref{pd_prop_eq_5}) we can infer that pole-dipole particles move according to 
\begin{eqnarray}
\frac{d}{dt} \overline{T}^{i0} &=&\stackrel{\{\,\}}{R}{}^{i}{}_{jkl} \overline{\Delta}^{klj}-{\stackrel{\{\,\}}{\Gamma}}_{kj}{}^{i} \overline{T}^{(kj)},  \label{sp_riem_pd_prop_eq_1}\\
v^{a} \overline{T}^{i0}&=& \overline{T}^{ia},  \label{sp_riem_pd_prop_eq_2} \\
\frac{d}{dt}\overline{\Delta}^{kl0} &=&-{\stackrel{\{\,\}}{\Gamma}}_{mj}{}^{k} \overline{\Delta}^{mlj}-{\stackrel{\{\,\}}{\Gamma}}_{mj}{}^{l}\overline{\Delta}^{k(mj)} -\overline{T}^{lk}+\overline{t}^{kl}, \label{sp_riem_pd_prop_eq_4} \\
0 &=&\overline{\Delta}^{kla}-v^{a}\overline{\Delta}^{kl0}-\overline{T}^{alk}+\overline{t}^{akl}.  \label{sp_riem_pd_prop_eq_5}
\end{eqnarray}

\paragraph{Single-pole particles}

Further restriction to single-pole particles with vanishing intrinsic hypermomentum $\Delta^{abc}$ brings the set (\ref{sp_riem_pd_prop_eq_1})-(\ref{sp_riem_pd_prop_eq_5}) into the form
\begin{eqnarray}
\frac{d}{dt}\overline{T}^{i0} &=&-{\stackrel{\{\,\}}{\Gamma}}_{kj}{}^{i}\overline{T}^{(kj)}, \nonumber \\
\overline{T}^{ia} &=&v^{a}\overline{T}^{i0},   \nonumber \\
\overline{T}^{lk} &=&\overline{t}^{kl}.  \nonumber 
\end{eqnarray}
As we have already shown these equations lead to the geodesic equation. Hence, within our general formalism we can quickly reproduce the standard result of general relativity.

\subsection{Open problems}

The results obtained in this work are valid for a wide class of extended gravitational theories that are naturally embedded into the framework of metric-affine gravity. However, our study is not exhaustive in many important aspects, and at this stage there remain several interesting open questions related to the multipole expansion of the equations of motion of test particles in alternative gravity theories. 

\subsubsection{Invariant definition of moments}

As we have already mentioned in previous sections, the definition of the integrated moments of the matter currents in the multipole formalism is to a certain extent ambiguous. This is related to the index positions in the integrand expression and to the nonconstancy of the metric which is used to lower and raise the indices. In view of this problem, we decided to present the full set of propagation equations for two different choices of the integrated moments, defined in equation (\ref{int_moments_definitions}) and (\ref{DY_int_moments_definitions}), respectively. Thereby one covers the definitions which have been discussed most frequently in the literature. Although we clearly favor the definition with mixed indices (\ref{DY_int_moments_definitions}), for the formal reasons given in section \ref{sec_propa_eq_alt}, even other index positions than the ones investigated in the present work are imaginable. Such an ambiguity in the definition of the integrated moments motivates the search for an invariant formulation. The corresponding program was already carried out in several works within a general relativistic context \cite{Tulczyjew:1959,Tulczyjew:1962,Dixon:1964,Beiglboeck:1967,Madore:1969}. Within an alternative gravity theory like metric-affine gravity, which is no longer a purely metric theory but has a richer geometrical structure, a detailed investigation is needed in order to generalize the concepts linked to such an invariant formulation.

\subsubsection{Supplementary conditions}

Previous analyses \cite{Frenkel:1926,Corinaldesi:Papapetrou:1951,Pirani:1956,Tulczyjew:1959} in metric theories of gravitation have shown, that even at the dipole level, supplementary conditions are needed in order to obtain a closed set of propagation equations. Indeed, let us recall the propagation equations in the Riemann-Cartan spacetime, for example. The four equations (\ref{Putau}) are sufficient to find the four coordinates of a position of a particle on its trajectory. However, the system (\ref{dottau}) contains only three independent equations, and this is not sufficient to determine six components of the spin. As a result, the supplementary conditions are usually imposed on the spin of the test particles in order to make number of the equations equal to the number of unknown variables. The imposition of an additional supplementary condition comes with some assumptions about the  physical nature of the particles under consideration, and there is no unique prescription how to do it. Even within the context of general relativity, a number of competing conditions exists. Furthermore, there seems to be no consensus on which of the supplementary conditions is the most physical one. In the context of alternative gravity theories the spectrum of possible supplementary conditions is greatly enhanced. This fact can be ascribed to the additional degrees of freedom within such theories, in particular, regarding the matter variables describing the internal structure of particles. Although there exist several studies of such supplementary conditions in the literature, most of them in the context of Einstein-Cartan and Poincar\'e gauge theory, a systematic and up-to-date analysis in the context of metric-affine gravity is still an outstanding task. We only note that an ultimate judgment over the correct choice of a supplementary condition can only be made with the help of an experiment. 

\subsubsection{Propagation equations involving higher moments}

If we take into account previous results in Einstein's theory \cite{Taub:1965}, it is to be expected that the role of supplementary conditions is even aggravated at higher orders of approximation. Of course this is due to the fact that at higher orders we need an even more detailed description of the internal dynamics of the test particles. Nevertheless, the study of higher orders of the propagation equations, beyond the pole-dipole level, will be of great interest in the context of radiation phenomena. In particular, we expect that such studies will shed light on our understanding of the new field strengths of metric-affine gravity, i.e., torsion and nonmetricity, which have no counterpart in the classical theory gravitation, namely general relativity. 

\subsubsection{Relation to other approximation schemes}

From a more formal standpoint, we can also ask about the compatibility with other approximation schemes which were employed in the context of gravitational theories. The most prominent examples being the post-Minkowskian and post-Newtonian approximation. Since these approximation schemes, in their full generality, are still under construction in the context of metric-affine gravity, a systematic comparison with the results obtained within a multipole scheme appears to be a long term project. 

\medskip 

To sum up, the study of the propagation equations of deformable test particles with the help of a multipole approximation scheme is a very rich field of research. In the context of alternative gravity theories this field is still in its infancy. Apart from the first steps undertaken in this work a number of open problems remain; we intend to attack these in future works.

\begin{acknowledgments}
The authors are grateful to F.W.\ Hehl (Univ.\ Cologne) for stimulating discussions and constructive criticism. Y.N.O.\ was supported by the Deutsche Forschungsgemeinschaft (Bonn) with the grant HE 528/21-1. D.P.\ acknowledges the support by {\O}.\ Elgar{\o}y (Univ.\ Oslo) and the Research Council of Norway under the project number 162830. 
\end{acknowledgments}

\appendix

\section{General conventions and notations}\label{general_conventions_sec}

In the theory of metric-affine gravity, the gravitational field is described by the three basic variables: the metric $g_{\alpha\beta}$, the coframe $\vartheta^\alpha$, and the linear connection $\Gamma_\alpha{}^\beta$. The Latin indices $i,j,\dots$ are used for local holonomic spacetime coordinates and the Greek indices $\alpha,\beta,\dots$ label (co)frame components. The vector basis dual to the frame 1-forms $\vartheta^\alpha$ is denoted by $e_\alpha$ and they satisfy $e_\alpha\rfloor\vartheta^\beta=\delta^\beta_\alpha$. Here $\rfloor$ denotes the interior product (contraction) of a vector with an exterior form. Using local coordinates $x^i$, we have $\vartheta^\alpha=h^\alpha_idx^i$ and $e_\alpha=h^i_\alpha\partial_i$. All objects and equations that carry the local Lorentz indices can be recast into their counterparts with the coordinate indices with the help of the contraction with the components of the tetrads, $h^\alpha_i$ and $h^i_\alpha$.

\subsection{Geometrical objects}\label{geometrical_objects_subsec}

The geometry of MAG is described by the {\it curvature} 2-form $R_{\alpha}{}^{\beta} := d\Gamma_{\alpha}{}^{\beta}+\Gamma_{\gamma}{}^{\beta}\wedge\Gamma_{\alpha}{}^{\gamma}$, the {\it nonmetricity} 1-form $Q_{\alpha\beta}:=-Dg_{\alpha\beta}$, and the {\it torsion} 2-form $T^{\alpha} :=D\vartheta^{\alpha}$ which are the gravitational field strengths for linear connection $\Gamma_{\alpha}{}^{\beta}$, metric $g_{\alpha\beta}$, and coframe $\vartheta^{\alpha}$, respectively. 

It is convenient to define a 1-form tensor-valued difference of the Riemannian (Christoffel) connection and the general linear connection:
\begin{equation}
N_\alpha{}^\beta := {\stackrel{\{\,\}}{\Gamma}}_\alpha{}^\beta- \Gamma_\alpha{}^\beta.\label{K}
\end{equation}
This quantity is known as {\it distorsion} 1-form. In particular, the torsion is recovered from it as $T^\alpha = - N_\beta{}^\alpha \wedge\vartheta^\beta$, whereas the nonmetricity arises as $Q_{\alpha\beta}= -2 N_{(\alpha\beta)}$. The corresponding curvature 2-forms are related via 
\begin{equation}\label{RR} 
R_\alpha{}^\beta = {\stackrel{\{\,\}}{R}}_\alpha{}^\beta - {\stackrel{\{\,\}}{D}}N_\alpha{}^\beta + N_\gamma{}^\beta\wedge N_\alpha{}^\gamma.
\end{equation}

\subsection{Physical objects}\label{physical_objects_subsec}

The sources of the metric-affine gravitational field are the 3-forms of the canonical energy-momentum and hypermomentum. They are defined by the variational derivatives of the material Lagrangian 4-form $L_{\rm mat}$, respectively:
\begin{eqnarray}
\Sigma_\alpha &=& {\frac {\delta L_{\rm mat}}{\delta\vartheta^\alpha}},\label{Sig}\\
\Delta^\alpha{}_\beta &=& {\frac {\delta L_{\rm mat}}{\delta\Gamma_\alpha{}^\beta}}.
\label{Del}
\end{eqnarray}
The Lagrangian $L_{\rm mat}$ also depends on some matter fields $\psi$, but this is irrelevant for the current discussion. 

\subsection{Components}\label{components_subsec}

When the local coordinates $x^i$ are chosen, we can write all the geometrical and physical quantities explicitly in terms of their components:
\begin{eqnarray}
\vartheta^\alpha &=& h^\alpha_i\,dx^i,\label{vtac}\\
\Gamma_\alpha{}^\beta &=& \Gamma_{i\alpha}{}^\beta\,dx^i,\label{gamc}\\
N_\alpha{}^\beta &=& N_{i\alpha}{}^\beta\,dx^i,\label{nc}\\ \label{rc}
R_\alpha{}^\beta &=& {\frac 12}\,R_{ij\alpha}{}^\beta\,dx^i\wedge dx^j,\\
\Sigma_\alpha &=& T_\alpha{}^i\,\partial_i\rfloor\eta,\label{sigc}\\
\Delta^\alpha{}_\beta &=& \Delta^\alpha{}_\beta{}^i\,\partial_i \rfloor\eta.\label{delc}
\end{eqnarray}
Here $\eta$ is the volume 4-form. Writing the Lagrangian form as $L_{\rm mat} = {\cal L}_{\rm mat}\,dx^0\wedge dx^1\wedge dx^2\wedge dx^3$, we can recast the definitions (\ref{Sig}) and (\ref{Del}) as follows:
\begin{eqnarray}
\widetilde{T}_\alpha{}^i &=& {\frac {\delta {\cal L}_{\rm mat}}{\delta h^\alpha_i}},\label{Sigi}\\
\widetilde{\Delta}^\alpha{}_\beta{}^i &=& {\frac {\delta {\cal L}_{\rm mat}}{\delta\Gamma_{i\alpha} {}^\beta}}.\label{Deli}
\end{eqnarray}

\section{Dimensions \& Symbols}\label{dimension_acronyms_app}

In order to fix our notation, we provide some tables with definitions in this appendix. The dimensions of the different quantities appearing throughout the work are displayed in table \ref{tab_dimensions}. Table \ref{tab_symbols} contains a list with symbols used throughout the text.

\begin{table}
\caption{\label{tab_dimensions}Dimensions of the quantities within this work.}
\begin{ruledtabular}
\begin{tabular}{cl}
Dimension (SI)&Symbol\\
\hline
&\\
\hline
Geometrical quantities&\\
\hline
1&$g_{\alpha\beta}$, $\delta _{\alpha \beta}$, $g_{ij}$, $\sqrt{-g}$, $h^\alpha_i$, $\Gamma_\alpha{}^\beta$, $N_\alpha{}^\beta$, $K_\alpha{}^\beta$, $R_\alpha{}^\beta$, $Q_{\alpha\beta}$, $Q$, $\ell^\alpha{}_\beta$\\
m&$x^{i}$, $dx^i$, $ds$, $\delta x^i$, $Y^a$, $\vartheta^\alpha$, $T^\alpha$\\
m$^{-1}$& $e_\alpha$, $\Gamma_{i\alpha}{}^\beta$, $N_{i\alpha}{}^\beta$, $K_{i\alpha}{}^\beta$, $Q_{i\alpha\beta}$, $Q_i$, $S_{ij}{}^k$\\
m$^{-2}$ & $R_{ij\alpha}{}^\beta$, $f_{ij}$ \\
m$^4$ & $\eta$\\
&\\
\hline
Matter quantities&\\
\hline
1& $u^{\alpha}$, $v^a$, $\rho^a{}_b$, $\psi$\\
kg\,m$^2/$s &  $h$ (Planck constant), $L$, $L_{\rm mat}$, $L_{\rm tot}$, $\Delta^\alpha{}_\beta$, $\tau_{\alpha\beta}$, $\sigma^{\alpha\beta}$, $m^{\alpha\beta}$, $M^{\alpha\beta}$, $H^\alpha{}_\beta$, $E^\alpha{}_\beta$, $\underline{\Delta}^i{}_j{}^k$, $\overline{\Delta}^{ijk}$, $\underline{T}^i{}_j{}^k$, $\underline{t}^{ij}{}_k$,  \\
&$\overline{T}^{ijk}$, $\overline{t}^{ijk}$, $\underline{Y}^k{}_l$, $\underline{\tau}^k{}_l$, $\underline{L}^k{}_l$, $\underline{\Lambda}^k{}_l$, $\underline{\tau}^k{}_l{}^j$, $\underline{Z}^k$, $\underline{Z}$ \\
kg\,m$/$s & $H_\alpha$, $E_\alpha$, $\Sigma_\alpha$, $\underline{T}_i{}^k$, $\underline{t}^i{}_j$, $\overline{T}^{ij}$, $\overline{t}^{ij}$, $\underline{P}_i$, ${\cal P}_i$, $m$ \\
kg$/$(m s)& $\Delta^\alpha{}_\beta{}^i$ \\
kg$/$(m$^2$s)& $T_\alpha{}^i$, ${\cal L}_{\rm mat}$\\
&\\
\hline
Operators&\\
\hline
1&$d$, $D$ \\
m$^{-1}$& $\partial_i$, $\nabla_i$, $\nabla_v$, ${\stackrel{\{\,\}}{\hbox{\L}}}_\xi$\\
\end{tabular}
\end{ruledtabular}
\end{table}

\begin{table}
\caption{\label{tab_symbols}Directory of symbols.} 
\begin{ruledtabular}
\begin{tabular}{lllc}
Symbol&&Explanation&Form degree\\
\hline
{\tiny Differential form} & {\tiny Component}&&\\
\multicolumn{2}{c}{{\tiny notation}}         &&\\
&&&\\
\hline
Geometrical quantities&&&\\
\hline
$g_{\alpha \beta}$& $g_{a b}$ & Metric& 0\\
                  & $g$ & Determinant of the metric&0\\
$\eta$&&Volume form &4\\      
$\vartheta^{\alpha}$& &Coframe&1\\
$T^{\alpha}$& $S_{ij}{}^k$ &Torsion&2\\
$e_{\alpha}$& &Vector basis&0\\
$Q_{\alpha \beta}$& $Q_{ijk}$ &Nonmetricity (Weyl 1-form denoted by $Q=Q_i dx^i$)&1\\
$R_{\alpha}{}^\beta$, $\stackrel{\{ \}}{R}_{\alpha}{}^\beta$& $R_{ijk}{}^l$, $\stackrel{\{\,\}}{R}_{ijk}{}^l$&General curvature, Riemannian curvature&2\\
&$\widehat{R}_{ijkl}$&Curvature ``object'' [defined in eq.\ (\ref{Rhat})]&0\\
$\Gamma_\alpha{}^\beta$, ${\stackrel{\{\,\}}{\Gamma}}_\alpha{}^\beta$& $\Gamma_{ij}{}^k$, $\stackrel{\{\,\}}{\Gamma}_{ij}{}^k$&Linear connection, Riemannian (Christoffel) connection&1\\
$N_\alpha{}^\beta$& $N_{ij}{}^k$ &Distorsion&1\\
$K_\alpha{}^\beta$& $K_{ij}{}^k$ &Contortion (antisymmetric part of the distorsion)&1\\ 
&$Y^a$&Worldline within the worldtube of the test particle& 0 \\
&$u^a$&Velocity along the worldline $Y^a$ of the particle&0\\
&&&\\
\hline
Matter quantities&&&\\
\hline
$L_{\rm tot}, L, L_{\rm mat}$& &Total, gravitational, matter Lagrangian & 4\\
$\sigma^{\alpha \beta}$& $t^{ij}$&Symmetric energy-momentum current& 4\\
$\Sigma_{\alpha}$& $T_i{}^j$ &Canonical energy-momentum current& 3\\
$\Delta^\alpha{}_{\beta}$& $\Delta^i{}_j{}^k$ &Hypermomentum current& 3\\
&$\overline{\underline{\Delta}}^{b_{1}\cdots b_{n}ijk}$&n-th integrated moment of the hypermomentum&0\\
&$\overline{\underline{T}}^{b_{1}\cdots b_{n}ij}$&n-th integrated moment of the canonical energy-mom.\ &0\\
&$\overline{\underline{t}}^{b_{1}\cdots b_{n}ij}$&n-th integrated moment of the symmetric energy-mom.\ &0\\
&$\overline{\underline{P}}_i$&Generalized integrated momentum &0\\
&$\overline{\underline{L}}^{a b}$&Generalized integrated orbital momentum&0\\
&$\overline{\underline{\Lambda}}^{a b}$&Antisymmetric part of the gen.\ int.\ orbital momentum&0\\
&$\overline{\underline{Y}}^{a b}$&Generalized integrated hypermomentum&0\\
&$\overline{\underline{Z}}$, $\overline{\underline{Z}}^k$&Dilaton part, i.e.\ the trace, of the generalized int.\ hypermomentum&0\\
$\tau_{\alpha \beta}$& $\tau_{ij}{}^k$ &Spin current (antisymmetric part of the hypermomentum current)& 3\\
&$J_A$&Placeholder for the density of a matter current (e.g.\ $\widetilde{\Delta }^{klj},$ $\widetilde{T}^{ij},$ or $\widetilde{t}^{kl}$)& 0\\
$$&$\psi$&Placeholder for a general matter field&0\\
$$&${\cal P}_i$&Generalized total 4-momentum [defined in eq.\ (\ref{Ptot})]&0\\
$$&$$&&\\
\hline
Operators&&&\\
\hline
$D$, $\stackrel{\{\,\}}{D}$ & $\nabla_i$, $\stackrel{\{\,\}}{\nabla}_i$ &Covariant (exterior) derivative, Riemannian covariant (exterior) derivative& $n\rightarrow n+1$\\
 & $\nabla_v$ &Convective covariant derivative (see, e.g., eq.\ (\ref{implicit_def_fluid_deriv}))& $n\rightarrow n+1$\\

$d$ & $,i$ &Exterior/partial derivative& $n\rightarrow n+1$\\
${\stackrel{\{\,\}}{\hbox{\L}}}_\xi$ & & Riemannian covariant Lie derivative & $n\rightarrow n$\\
&$\rho^a{}_b$&Spatial projector (equals the convective part, denoted by $\stackrel{(c)}{}$) &0\\
&&&\\
\hline
Accents&&&\\
\hline
         &``$\stackrel{(c)}{}$''& Denotes the convective part of an object&\\
Tilde    &``$\widetilde{\phantom{pen}}$''&Denotes the density of an object&\\
Overline &``$\overline{\phantom{pen}}$''&Denotes integrated version of a density based on upper-index convention&\\
Underline&``$\underline{\phantom{pen}}$''&Denotes integrated version of a density based on lower-index convention&\\
\end{tabular}
\end{ruledtabular}
\end{table}

\bibliographystyle{unsrtnat}
\bibliography{eom_bibliography}

\end{document}